\documentclass[12pt]{article}

\usepackage[totalwidth=460truept,totalheight=600truept]{geometry}
\usepackage{latexsym,graphicx,amsfonts,amssymb,amsmath}
\usepackage[hypertexnames=false,hidelinks]{hyperref}

\def\theequation{\arabic{section}.\arabic{equation}}

\renewcommand{\theequation}{\thesection.\arabic{equation}}
\global\arraycolsep=1truept
\renewcommand{\theequation}{\arabic{section}.\arabic{equation}}

\begin{document}

\null

\vskip1truecm

\begin{center}
{\huge \textbf{Predictions of Quantum Gravity}}

\vskip.3truecm

{\huge \textbf{in Inflationary Cosmology:}}

\vskip.3truecm

{\huge \textbf{Effects of the Weyl-squared Term}}

\vskip1truecm

\textsl{Damiano Anselmi}$^{1,2,a}$\textsl{, Eugenio Bianchi}$^{3,4,b}$%
\textsl{\ and Marco Piva}$^{5,c}$

\vskip .1truecm

$^{1}${\scriptsize \textit{Dipartimento di Fisica \textquotedblleft Enrico
Fermi\textquotedblright , Universit\`{a} di Pisa, Largo B. Pontecorvo 3,
56127 Pisa, Italy}}

$^{2}${\scriptsize \textit{INFN, Sezione di Pisa, Largo B. Pontecorvo 3,
56127 Pisa, Italy}}

$^{3}${\scriptsize \textit{Institute for Gravitation and the Cosmos, The
Pennsylvania State University, University Park, Pennsylvania 16802, USA}}

$^{4}${\scriptsize \textit{Department of Physics, The Pennsylvania State
University, University Park, Pennsylvania 16802, USA}}

$^{5}${\scriptsize \textit{National Institute of Chemical Physics and
Biophysics, R\"{a}vala 10, Tallinn 10143, Estonia}}

$^{a}${\footnotesize damiano.anselmi@unipi.it, }$^{b}${\footnotesize %
ebianchi@gravity.psu.edu, }$^{c}${\footnotesize marco.piva@kbfi.ee}

\vskip 1.5truecm

\textbf{Abstract}
\end{center}

We derive the predictions of quantum gravity with fakeons on the amplitudes
and spectral indices of the scalar and tensor fluctuations in inflationary
cosmology. The action is $R+R^{2}$ plus the Weyl-squared term. The ghost is
eliminated by turning it into a fakeon, that is to say a purely virtual
particle. We work to the next-to-leading order of the expansion around the
de Sitter background. The consistency of the approach puts a lower bound ($%
m_{\chi }>m_{\phi }/4$) on the mass $m_{\chi }$ of the fakeon with respect
to the mass $m_{\phi }$ of the inflaton. The tensor-to-scalar ratio $r$ is
predicted within less than an order of magnitude ($4/3<N^{2}r<12$ to the
leading order in the number of $e$-foldings $N$). Moreover, the relation $%
r\simeq -8n_{T}$ is not affected by the Weyl-squared term. No vector and no
other scalar/tensor degree of freedom is present.

\vfill\eject

\section{Introduction}

\label{intro}\setcounter{equation}{0}

Inflation is a theory of accelerated expansion of the early universe \cite%
{englert,starobinsky,kazanas,sato,guth,linde,steinhardt,linde2}, which
accounts for the origin of the present large-scale structure. It explains
the approximate isotropy of the cosmic microwave background radiation and
allows us to study the quantum fluctuations as sources of the cosmological
perturbations that seed the formation of the structures of the cosmos \cite%
{mukh,mukh2,hawk,guth2,staro2,bardeen,mukh3}. It also provides a rich
environment where we can develop knowledge that might allow us to establish
a nontrivial connection between high-energy physics and the physics of large
scales.

Inflationary cosmology is often studied with the help of a matter field that
drives the expansion by rolling down a potential $V(\phi )$ (for reviews,
see \cite{weincosmo,reviews}). Alternatively, gravity itself can drive the
expansion, as in the Starobinsky $R+R^{2}$ model \cite{starobinsky} and the $%
f(R)$ theories \cite{defelice,otherfR}. The predictions end up depending
strongly on the model, specifically the choices of $V(\phi )$ and $f(R)$. In
single-field slow-roll inflation, potentials with a plateau lead to a scalar
power spectrum that is compatible with current observations \cite%
{encicl,Planck18}.

In particular, the Starobinsky $R+R^{2}$ model works well at the
phenomenological level. However, once $R^{2}$ is introduced, it is hard to
justify why the square $C_{\mu \nu \rho \sigma }C^{\mu \nu \rho \sigma }$ of
the Weyl tensor $C_{\mu \nu \rho \sigma }$ is not included as well, since it
has the same dimension in units of mass. We can spare the other quadratic
combinations, such as $R_{\mu \nu }R^{\mu \nu }$ and $R_{\mu \nu \rho \sigma
}R^{\mu \nu \rho \sigma }$, since they are related to $R^{2}$ and $C_{\mu
\nu \rho \sigma }C^{\mu \nu \rho \sigma }$ by algebraic identities and the
Gauss-Bonnet theorem. Thus, we are lead to consider the action 
\begin{equation}
S=-\frac{M_{\text{Pl}}^{2}}{16\pi }\int \mathrm{d}^{4}x\sqrt{-g}\left(
R+\alpha \hspace{0.01in}R^{2}+\beta \hspace{0.01in}C_{\mu \nu \rho \sigma
}C^{\mu \nu \rho \sigma }\right) ,  \label{theo}
\end{equation}%
which we briefly refer to as \textquotedblleft $R+R^{2}+C^{2}$
theory\textquotedblright . The trouble with (\ref{theo}) is that the $C^{2}$
term is normally responsible for the presence of ghosts. Immediate ways out
are to expand the physical quantities in powers of $\beta $ \cite%
{weinbergeffe}, which is equivalent to assume that the ghosts are very
heavy, and/or restrict to situations where the ghosts are short-lived. This
approach amounts to \textquotedblleft living with ghosts\textquotedblright\ 
\cite{hawking}, but does not eliminate the problem.

If we want to work with the $R+R^{2}+C^{2}$ theory, we must explain how to
treat $C^{2}$ in order to remove the ghosts, at least perturbatively and at
the level of the cosmological perturbations. Here we use the procedure of
eliminating them in favor of purely virtual particles \cite{LWgrav,fakeons}.
This procedure originates in high-energy physics, where the requirements of
locality, renormalizability and unitarity result in consistency contraints
on perturbative quantum field theory.

The simplest way to think of the idea is as follows. A normal particle can
be real or virtual, depending on whether it is observed or not. As far as we
know, a particle that is always real does not exist. What about a particle
that is always virtual and can never become real? We can think of it as a
purely virtual quantum \cite{wheelerons} or a fake particle, i.e., a
particle that mediates interactions among other particles, but is invisible
to our detectors. And by that we mean invisible in principle, not just in
practice.

Perturbative quantum gravity can be formulated as a unitary theory of
scattering if the action (\ref{theo}) is quantized in a new way \cite{LWgrav}%
, by eliminating the would-be ghost in favor of a purely virtual particle,
called \textit{fakeon}\ \cite{fakeons}. In the expansion around flat space,
the fakeon is introduced by replacing the Feynman $i\epsilon $ prescription
(for a pole of the free propagator) with an alternative prescription that
allows us to project the corresponding degree of freedom away consistently
with the optical theorem. This means that the loop corrections are unable to
resuscitate the degree of freedom back. Moreover, the prescription is
compatible with renormalizability \cite{LWgrav,fakeons}. A fakeon mediates
interactions, but does not belong to the spectrum of asymptotic states. In
this sense it is a \textquotedblleft fake degree of
freedom.\textquotedblright\ Note that it removes a ghost at the fundamental
level, without advocating its irrelevance for practical purposes.
Incidentally, the calculations of Feynman diagrams with the fakeon
prescription in quantum gravity \cite{UVQG,absograv} are not harder than
analogous calculations for the standard model.

Nevertheless, quantum field theory is formulated perturbatively, commonly
around flat space. To study inflation and cosmology it is necessary to work
on nontrivial backgrounds. This raises the issue of understanding purely
virtual quanta in curved space. A simplification comes from the fact that in
cosmology we do not need to go as far as computing loop corrections, as
argued in ref. \cite{weinberg}, although we have to study the quantum
fluctuations. In this paper, we show that we can work with the classical
limit of the fakeon prescription/projection, which amounts to taking the
average of the retarded and advanced potentials as Green function $G_{\text{f%
}}$ for the fake particles \cite{causalityQG},%
\begin{equation}
G_{\text{f}}=\frac{1}{2}\left( G_{\text{Ret}}+G_{\text{Adv}}\right) ,
\label{gf}
\end{equation}%
combined with a certain wealth of knowledge on how to use this formula and
interpret its consequences. Note that the quantum fakeon prescription cannot
be inferred from (\ref{gf}), because (\ref{gf}) is not a good propagator in
Feynman diagrams \cite{wheelerons}.

As said, the predictions of the popular models of inflation are model
dependent. On the other hand, in high-energy physics the constraints of
locality, unitarity and renormalizability leave room for a limited class of
interactions, scalar potentials, and so on, to the extent that the theory of
quantum gravity emerging from the idea of fake particle is essentially
unique (when matter is switched off) and contains just two independent
parameters more than Einstein gravity. They can be identified as the masses $%
m_{\phi }$ and $m_{\chi }$ of a scalar field $\phi $ (the inflaton) and a
spin-2 fakeon $\chi _{\mu \nu }$. The triplet graviton-scalar-fakeon
exhausts the set of degrees of freedom of the theory. From the cosmological
point of view the physical modes are the usual curvature perturbation $%
\mathcal{R}$ and the tensor fluctuations. The extra degrees of freedom are
turned into fake ones and projected away. In particular, no vector
fluctuations, or additional scalar and tensor fluctuations survive.

We show that the consistency of the picture in curved space leads to a lower
bound $m_{\chi }>m_{\phi }/4$ on the mass $m_{\chi }$ of the fakeon with
respect to the mass $m_{\phi }$ of the inflaton. To the next-to-leading
order, the amplitude $A_{\mathcal{R}}$ and the spectral index $n_{\mathcal{R}%
}-1$ of the scalar fluctuations depend only on $m_{\phi }$ (and the number $%
N $ of $e$-foldings). Instead, the amplitude $A_{T}$ and the spectral index $%
n_{T}$\ of the tensor fluctuations do depend on $m_{\chi }$. The bound $%
m_{\chi }>m_{\phi }/4$ narrows the window of allowed values of $n_{T}$ and
the tensor-to-scalar ratio $r=A_{T}/A_{\mathcal{R}}$ to less than one order
of magnitude and makes the predictions quite precise, even before knowing
the actual values of $m_{\phi }$ and $m_{\chi }$.

Inflationary cosmology in higher-derivative gravity with ghosts have been
studied in refs. \cite{deruelle,salvio}. Typically, the ghost sector is
quantized by means of negative norms. Extra spectra are predicted, which may
or may not be suppressed on superhorizon scales. Inflation has been
considered in nonlocal theories of gravity as well \cite{modesto}, where the
classical action contains infinitely many free parameters. The cosmological
perturbations in those scenarios have been studied in \cite{staro}.

The gain achieved by means of fakeons is that no ghosts are present and the
number of independent parameters is kept to a minimum. Whenever there is an
overlap, we find agreement with the results derived in the other approaches.
This occurs, for example, when $H/m_{\chi }$ or $m_{\phi }/m_{\chi }$ are
sufficiently small to suppress the effects of the fakeons in our theory and
the effects of the ghosts in the theories of refs. \cite{salvio}, where $H$
is the value of the Hubble parameter during inflation. Even when $H$ or $%
m_{\chi }$ are not large, we can still relate some results, due to the
universality of the low-energy expansion. For example, we can do so for any
quantity that has a convergent, resummable expansion for small $H/m_{\chi }$
or $m_{\phi }/m_{\chi }$. In the limit $m_{\chi }/m_{\phi }\rightarrow
\infty $, the results we find agree with those of the theory $R+R^{2}$ \cite%
{defelice,mukhanov}.

We make the calculations in two frameworks and show that the final results
match. In the first approach, which we call \textit{inflaton framework}, the
scalar field $\phi $ is introduced explicitly to eliminate the $R^{2}$ term,
while the $C^{2}$ term is unmodified. The scalar potential coincides with
the Starobinsky one. In the second approach, which we call \textit{geometric
framework}, both $R^{2}$ and $C^{2}$ are present. The $C^{2}$ term does not
affect the FLRW metric, so in both approaches the background metric
coincides with the one of the Starobinsky theory. The differences arise in
the action of the fluctuations over the background. The map relating the two
frameworks is a field-dependent conformal transformation, combined with a
time reparametrization. A third formulation, where the scalar $\phi $ and a
spin-2 fakeon $\chi _{\mu \nu }$ are introduced explicitly in order to
eliminate both higher-derivative terms $R^{2}$ and $C^{2}$ is also available 
\cite{absograv}, but will not be studied here.

The paper is organized as follows. In section \ref{theory}, we briefly
review the formulation of quantum gravity with fakeons and present the two
frameworks just mentioned. In section \ref{inflaton}, we study the tensor
and scalar fluctuations in the inflaton framework. The fakeon projection,
which allows us to make sense of the term $C^{2}$, is briefly introduced in
section \ref{theory} and discussed in detail in section \ref{proj}. In
section \ref{geometric}, we make the calculations in the geometric
framework. In section \ref{vector}, we study the vector fluctuations and
show that they are projected away altogether at the quadratic level. Section %
\ref{predictions} contains the summary of our predictions and section \ref%
{conclusions} contains the conclusions. In appendix \ref{matching}, we
derive the map relating the inflaton framework to the geometric framework
and show that the results agree. In appendix \ref{appe} we show that the
curvature perturbation $\mathcal{R}$ can be considered constant on
superhorizon scales for adiabatic fluctuations of the energy-momentum tensor.

\section{Quantum gravity with fakeons}

\label{theory}\setcounter{equation}{0}

In this section we introduce the theory and the two frameworks we are going
to work with. We begin by recalling a few basic features of the fakeons.
Being purely virtual quanta, they are particles that mediate interactions,
but do not belong to the physical spectrum of asymptotic states. Expanding
around flat space, they are introduced by means of a new quantization
prescription for the poles of the free propagators \cite{LWgrav},
alternative to the Feynman $i\epsilon $ prescription. The physical subspace $%
V$ is obtained by projecting the fake degrees of freedom away. The theory is
unitary in $V$, where the optical theorem holds. What makes the projection
consistent to all orders \cite{fakeons} is that the fakeon prescription does
not allow the loop corrections to resuscitate back the states that have been
projected away.

The prescription makes sense irrespective of the sign of the residue at the
pole of the propagator. Yet, it requires that the real part of the squared
mass be positive. Indeed, fakeons cannot cure tachyons, but only ghosts. The
no-tachyon condition is the main requirement we have to fulfill and its
analogue on nontrivial backgrounds is going to play an important role.

The projection must also be performed at the classical level. An action like
(\ref{theo}) is physically unacceptable as the classical limit of quantum
gravity, because it has undesirable solutions. Yet, (\ref{theo}) is the
starting point to formulate quantum gravity as a quantum field theory. It is
local and provides the Feynman rules that allow us (together with the
Feynman prescription for physical particles and the fakeon prescription for
fake particles), to calculate the loop diagrams and the $S$ matrix. An
action of this type is called \textquotedblleft interim\textquotedblright\
classical action\ \cite{causalityQG}.

The true classical action $S_{\text{class}}$ is obtained from the interim
classical action $S_{\text{inter}}$ by projecting the fake degrees of
freedom away. At the classical level, the projection is achieved by means of
the classical limit of the fakeon prescription. Precisely, $S_{\text{class}}$
is obtained by: ($i$) solving the field equations of the fakeons (derived
from $S_{\text{inter}}$) by means of the fakeon Green function; and ($ii$)
inserting the solutions back into $S_{\text{inter}}$. In the perturbative
expansion around flat space, the fakeon Green function is the arithmetic
average of the retarded and advanced potentials \cite{causalityQG}. We will
see that this piece of information is enough to derive the fakeon Green
function on nontrivial backgrounds.

The plan of the paper is to calculate the effects of inflationary cosmology
on the fluctuations of the cosmic microwave background radiation at the
quadratic level. Since we do not need to work out loop corrections, we can
quantize the projected action $S_{\text{class}}$, rather than projecting the
quantum version of $S_{\text{inter}}$. This simplification saves us a lot of
effort.

The good feature of $S_{\text{class}}$ is that it no longer contains the
fake degrees of freedom, by construction, so in principle it can be
quantized with the usual methods. The nontrivial counterpart is that $S_{%
\text{class}}$ is not fully local, due to the nonlocal remnants left by the
fakeon projection. Because of this, the quantization of $S_{\text{class}}$
is not as simple as usual, also taking into account that we must perform it
on a nontrivial background. However, in a variety of lucky cases, which
include those studied in this paper, it is possible to treat the nonlocal
sector of $S_{\text{class}}$ in a relatively simple way and extract physical
predictions with the procedure described above, either because the nonlocal
sector of $S_{\text{class}}$ does not affect the quantities we are
interested in, or because it affects them only at higher orders.

Summarizing, the simplest way to proceed, which we adopt in the paper, is as
follows. First, we work out the classical action $S_{\text{class}}$ of
quantum gravity, by projecting the interim action $S_{\text{inter}}$.
Second, we quantize $S_{\text{class}}$ with the usual methods, paying
special attention to the nonlocal sector, anticipating that in the end it
does not create too serious difficulties.

Now we give the interim\ classical actions $S_{\text{inter}}$ of quantum
gravity in the two approaches we study in the paper. The projection $S_{%
\text{inter}}\rightarrow S_{\text{class}}$ and the quantization of $S_{\text{%
class}}$ will be performed in the next sections, after expanding around the
de Sitter background.

The higher-derivative form of the interim\ classical action is%
\begin{equation}
S_{\text{geom}}(g,\Phi )=-\frac{M_{\text{Pl}}^{2}}{16\pi }\int \mathrm{d}%
^{4}x\sqrt{-g}\left[ R+\frac{1}{2m_{\chi }^{2}}C_{\mu \nu \rho \sigma
}C^{\mu \nu \rho \sigma }-\frac{R^{2}}{6m_{\phi }^{2}}\right] +S_{\mathfrak{m%
}}(g,\Phi ),  \label{SQG}
\end{equation}%
where $C_{\mu \nu \rho \sigma }$ denotes the Weyl tensor, $M_{\text{Pl}}=1/%
\sqrt{G}$ is the Planck mass, $\Phi $ are the matter fields and $S_{%
\mathfrak{m}}$ is the action of the matter sector. The no-tachyon condition
(i.e., the requirement that the free propagator around flat space does not
have tachyonic poles) determines the signs in front of $C_{\mu \nu \rho
\sigma }C^{\mu \nu \rho \sigma }$ and $R^{2}$.

The degrees of freedom of the gravitational sector are the graviton, a
scalar field $\phi $ of mass $m_{\phi }$ and a spin-2 fakeon $\chi _{\mu \nu
}$ of mass $m_{\chi }$. The reason why $\chi _{\mu \nu }$ must be quantized
as a fakeon is that the residue of the free propagator has the wrong sign at
the $\chi _{\mu \nu }$ pole, so the Feynman prescription would turn it into
a ghost, causing the violation of unitarity. On the other hand, $\phi $ can
be quantized either as a fakeon or a physical particle, because the residue
at the $\phi $ pole has the correct sign. In this paper, we assume that $%
\phi $ is a physical particle (the inflaton).

For simplicity, we have omitted the cosmological term in (\ref{SQG}). We
will do the same throughout the paper. Once it is included, the theory is
manifestly renormalizable, like Stelle's theory \cite{stelle}, because the
fakeon prescription does not modify the ultraviolet divergences \cite%
{LWgrav,fakeons}.

With the help of an auxiliary field $\varphi $, we can write $S_{\text{QG}}$
in the equivalent form%
\begin{eqnarray}
S_{\text{geom}} &=&-\frac{M_{\text{Pl}}^{2}}{16\pi }\int \mathrm{d}^{4}x%
\sqrt{-g}\left( R+\frac{1}{2m_{\chi }^{2}}C_{\mu \nu \rho \sigma }C^{\mu \nu
\rho \sigma }\right)  \notag \\
&&+\frac{M_{\text{Pl}}^{2}}{96\pi m_{\phi }^{2}}\int \mathrm{d}^{4}x\sqrt{-g}%
(2R-\varphi )\varphi +S_{\mathfrak{m}}(g,\Phi ).  \label{SQGmix}
\end{eqnarray}%
Making the Weyl transformation 
\begin{equation}
g_{\mu \nu }\rightarrow g_{\mu \nu }\mathrm{e}^{\hat{\kappa}\phi },\qquad
\phi =-\frac{1}{\hat{\kappa}}\ln \left( 1-\frac{\varphi }{3m_{\phi }^{2}}%
\right) ,  \label{weyl}
\end{equation}%
where $\hat{\kappa}=M_{\text{Pl}}^{-1}\sqrt{16\pi /3}$, we can diagonalize
the quadratic part and obtain the new action%
\begin{equation}
S_{\text{infl}}=-\frac{M_{\text{Pl}}^{2}}{16\pi }\int \mathrm{d}^{4}x\sqrt{-g%
}\left( R+\frac{1}{2m_{\chi }^{2}}C_{\mu \nu \rho \sigma }C^{\mu \nu \rho
\sigma }\right) +S_{\phi }(g,\phi )+S_{\mathfrak{m}}(g\mathrm{e}^{\hat{\kappa%
}\phi },\Phi ),  \label{sqgeq}
\end{equation}%
where 
\begin{equation}
S_{\phi }(g,\phi )=\frac{1}{2}\int \mathrm{d}^{4}x\sqrt{-g}\left( D_{\mu
}\phi D^{\mu }\phi -2V(\phi )\right)  \label{sphi}
\end{equation}%
and 
\begin{equation}
V(\phi )=\frac{m_{\phi }^{2}}{2\hat{\kappa}^{2}}\left( 1-\mathrm{e}^{\hat{%
\kappa}\phi }\right) ^{2}  \label{staropot}
\end{equation}%
is the Starobinsky potential. The action (\ref{sqgeq}) is not manifestly
renormalizable. In fact, it is as renormalizable as (\ref{SQG}) -- once the
cosmological term is reinstated --, because it is related to (\ref{SQG}) by
a (perturbative and nonderivative) field redefinition.

The geometric framework is defined by the interim actions (\ref{SQG}) or (%
\ref{SQGmix}), while the inflaton framework is defined by (\ref{sqgeq}). In
the rest of the paper, we switch the matter sector $S_{\mathfrak{m}}$ off.
If needed, its effects can be studied along the guidelines outlined in the
next sections. We do not review the details on the parametrizations of the
fluctuations and their transformations under diffeomorphisms, which are easy
to find in the literature (see for example \cite{reviews,defelice}).

\section[Inflaton framework]{Inflaton framework ($R+\hspace{0.01in}$scalar$%
\hspace{0.01in}+C^{2}$)}

\label{inflaton}\setcounter{equation}{0}

In this section, we study the tensor and scalar fluctuations in the inflaton
framework. The action is (\ref{sqgeq}), with the potential (\ref{staropot}).
The Friedmann equations are%
\begin{equation}
\frac{\ddot{a}}{a}-\frac{\dot{a}^{2}}{a^{2}}=-4\pi G\dot{\phi}^{2},\qquad 
\frac{\dot{a}^{2}}{a^{2}}=\frac{4\pi G}{3}\left( \dot{\phi}^{2}+2V(\phi
)\right) ,\qquad \ddot{\phi}+3\frac{\dot{a}}{a}\dot{\phi}=-V^{\prime }(\phi
),  \label{frie}
\end{equation}%
where $V^{\prime }(\phi )=\mathrm{d}V(\phi )/\mathrm{d}\phi $. We define the
usual quantities%
\begin{equation}
\varepsilon =-\frac{\dot{H}}{H^{2}},\qquad \eta =2\varepsilon -\frac{\dot{%
\varepsilon}}{2H\varepsilon },  \label{eta}
\end{equation}%
where $H=\dot{a}/a$ is the Hubble parameter.

The de Sitter limit is the one where $H$ is approximately constant. It is
easy to show that the constant value it tends to is $m_{\phi }/2$. Indeed, $%
\dot{H}\simeq 0$ in the first equation (\ref{frie}) gives $\dot{\phi}\simeq
0 $. On the other hand, if we insert $\dot{\phi}\simeq 0$ (and so $\ddot{\phi%
}\simeq 0$) in the third equation (\ref{frie}) we obtain $V^{\prime }(\phi
)\simeq 0$, which has two solutions: $\phi \simeq 0$ and $\phi \rightarrow
-\infty $. The first possibility gives the trivial case, since $\phi \simeq
0 $, $\dot{\phi}\simeq 0$ in the second equation (\ref{frie}) give $H\simeq
0 $. The second possibility is the right one, since $\phi \rightarrow
-\infty $, $\dot{\phi}\simeq 0$ in the second equation (\ref{frie}) give $%
H\simeq m_{\phi }/2$.

The expansion around the de Sitter background is an expansion in powers of $%
\sqrt{\varepsilon }$. This can be proved by studying the solution of the
equations (\ref{frie}) around the de Sitter metric. Leaving the details to
appendix \ref{matching}, here we just mention the properties that we need to
proceed. It is possible to show that $\eta =\mathcal{O}(\sqrt{\varepsilon })$
and 
\begin{equation}
\frac{\mathrm{d}^{n}\varepsilon }{\mathrm{d}t^{n}}=H^{n}\mathcal{O}%
(\varepsilon ^{\frac{n+2}{2}}).  \label{assum}
\end{equation}%
In other words, each time derivative raises the order by $\sqrt{\varepsilon }
$, so the expansion in powers of $\sqrt{\varepsilon }$ is also an expansion
of slow time dependence. Moreover, we have 
\begin{eqnarray}
H &=&\frac{m_{\phi }}{2}\left( 1-\frac{\sqrt{3\varepsilon }}{2}+\frac{%
7\varepsilon }{12}+\mathcal{O}(\varepsilon ^{3/2})\right) ,  \notag \\
\eta &=&-2\sqrt{\frac{\varepsilon }{3}}+\frac{13}{9}\varepsilon +\mathcal{O}%
(\varepsilon ^{3/2}),  \label{ah} \\
-aH\tau &=&1+\varepsilon +\mathcal{O}(\varepsilon ^{3/2}).  \notag
\end{eqnarray}%
(see formulas (\ref{hbar}), suppressing bars). The last line is the
expansion of $-aH\tau $, where $\tau $ is the conformal time, defined by%
\begin{equation}
\tau =-\int_{t}^{+\infty }\frac{\mathrm{d}t^{\prime }}{a(t^{\prime })},
\label{tau}
\end{equation}%
with the initial condition chosen to have $\tau =-1/(aH)$ in the de Sitter
limit $\varepsilon \rightarrow 0$.

\subsection{Tensor fluctuations}

\label{tensor}

To study the tensor fluctuations, it is convenient to parametrize the metric
as 
\begin{equation}
g_{\mu \nu }=\text{diag}(1,-a^{2},-a^{2},-a^{2})-2a^{2}\left( u\delta _{\mu
}^{1}\delta _{\nu }^{1}-u\delta _{\mu }^{2}\delta _{\nu }^{2}+v\delta _{\mu
}^{1}\delta _{\nu }^{2}+v\delta _{\mu }^{2}\delta _{\nu }^{1}\right) ,
\label{met}
\end{equation}%
where $u=u(t,z)$ and $v=v(t,z)$ are the graviton modes.

Let $u_{\mathbf{k}}(t)$ denote the Fourier transform of $u(t,z)$ with
respect to the coordinate $z$, where $\mathbf{k}$ is the space momentum. The
quadratic Lagrangian obtained from (\ref{sqgeq}) is 
\begin{equation}
(8\pi G)\frac{\mathcal{L}_{\text{t}}}{a^{3}}=\dot{u}^{2}-\frac{k^{2}}{a^{2}}%
u^{2}-\frac{1}{m_{\chi }^{2}}\left[ \ddot{u}^{2}-2\left( H^{2}-2\pi G\dot{%
\phi}^{2}+\frac{k^{2}}{a^{2}}\right) \dot{u}^{2}+\frac{k^{4}}{a^{4}}u^{2}%
\right] ,  \label{lt}
\end{equation}%
plus an identical contribution for $v$, where $k=|\mathbf{k}|$. To simplify
the notation, we understand that $u^{2}$ stands for $u_{-\mathbf{k}}u_{%
\mathbf{k}}$, $\dot{u}^{2}$ for $\dot{u}_{-\mathbf{k}}\dot{u}_{\mathbf{k}}$,
etc. We extend this convention to mixed products such as $u\dot{u}$, which
can be interpreted either as $u_{-\mathbf{k}}\dot{u}_{\mathbf{k}}$ or $\dot{u%
}_{-\mathbf{k}}u_{\mathbf{k}}$.

It is possible to eliminate the higher derivatives by considering the
extended Lagrangian 
\begin{equation}
\mathcal{L}_{\text{t}}^{\prime }=\mathcal{L}_{\text{t}}+\Delta \mathcal{L}_{%
\text{t}},  \label{ltp}
\end{equation}%
where 
\begin{equation}
(8\pi Gm_{\chi }^{2})\frac{\Delta \mathcal{L}_{\text{t}}}{a^{3}}=\left( S-%
\ddot{u}-f\dot{u}-hu\right) ^{2}.  \label{dlt}
\end{equation}%
Here $f(t)$, $h(t)$ are functions to be determined, and $S$, which may stand
for $S_{-\mathbf{k}}(t)$ or $S_{\mathbf{k}}(t)$, denotes an auxiliary field.
The equivalence of $\mathcal{L}_{\text{t}}^{\prime }$ and $\mathcal{L}_{%
\text{t}}$ is due to the fact that $\mathcal{L}_{\text{t}}^{\prime }=%
\mathcal{L}_{\text{t}}$ when $S$ is replaced by the solution of its own
field equation. The higher derivatives disappear in the sum $\mathcal{L}_{%
\text{t}}+\Delta \mathcal{L}_{\text{t}}$, because the term proportional to $%
\ddot{u}^{2}$ cancels out.

Next, we perform the field redefinitions 
\begin{equation}
u=U+\alpha V,\qquad \qquad S=V+\beta U,  \label{uU}
\end{equation}%
where $\alpha (t)$ and $\beta (t)$ are other functions to be determined. We
use the freedom to choose $f$, $h$, $\alpha $ and $\beta $ to write $%
\mathcal{L}_{\text{t}}^{\prime }$ in a convenient form, such that it
contains a unique, nonderivative term mixing $U$ and $V$. Specifically, we
reduce the Lagrangian $\mathcal{L}_{\text{t}}^{\prime }$ to the form 
\begin{equation}
\mathcal{L}_{\text{t}}^{\prime }=\mathcal{L}_{\text{t}}^{(U)}+\mathcal{L}_{%
\text{t}}^{(V)}+\mathcal{L}_{\text{t}}^{(UV)},  \label{LTp}
\end{equation}%
where 
\begin{eqnarray}
(8\pi G)\frac{\mathcal{L}_{\text{t}}^{(U)}}{a^{3}\gamma } &=&\dot{U}%
^{2}-\omega ^{2}U^{2},\qquad (8\pi Gm_{\chi }^{2}M^{2})\frac{\mathcal{L}_{%
\text{t}}^{(UV)}}{a^{3}\gamma }=2\sigma UV,  \notag \\
(8\pi GM^{4})\frac{\mathcal{L}_{\text{t}}^{(V)}}{a^{3}\gamma } &=&-\dot{V}%
^{2}+\Omega ^{2}V^{2},  \label{mix}
\end{eqnarray}%
and $\gamma $, $\omega ^{2}$, $\Omega ^{2}$ and $\sigma $ are other
functions of time, while $M$ is constant and has the dimension of a mass.
Since $\gamma $ is going to be positive, $V$ is the fakeon and $U$ is the
physical excitation, up to the mixing due to $\mathcal{L}_{\text{t}}^{(UV)}$.

The fakeon projection amounts to solving the $V$ field equations by means of
the fakeon prescription and inserting the solution back into $\mathcal{L}_{%
\text{t}}^{\prime }$. In all the cases considered here, this is achieved by
determining the solution $G_{\text{f}}(t,t^{\prime })$ of $\Sigma G_{\text{f}%
}(t,t^{\prime })=\delta (t-t^{\prime })$ as the arithmetic average of the
retarded and advanced potentials, where $\Sigma $ is an operator of the form%
\begin{equation}
\Sigma \equiv F_{2}(t)\frac{\mathrm{d}^{2}}{\mathrm{d}t^{2}}+F_{1}(t)\frac{%
\mathrm{d}}{\mathrm{d}t}+F_{0}(t),  \label{sigmagen}
\end{equation}%
$F_{i}(t)$ being functions of time. A certain detour allows us to get to the
results we need here without even knowing the explicit expression of $G_{%
\text{f}}(t,t^{\prime })$, which is derived in section \ref{proj}, where the
projection is discussed in detail.

If we take 
\begin{eqnarray}
\alpha &=&\frac{1}{M^{2}},\qquad \beta =M^{2}+m_{\chi
}^{2}+2H_{0}^{2},\qquad f=3H,  \notag \\
h &=&M^{2}+m_{\chi }^{2}+H_{0}^{2}+H^{2}\left( 1+\varepsilon \right) +\frac{%
k^{2}}{a^{2}},  \label{ab}
\end{eqnarray}%
where $H_{0}$ is a constant, we obtain the decomposition (\ref{LTp}) with 
\begin{eqnarray}
\gamma &=&1+2\frac{H_{0}^{2}}{m_{\chi }^{2}},\qquad \omega ^{2}=h-M^{2}-%
\frac{\sigma }{m_{\chi }^{2}}-m_{\chi }^{2}\gamma ,\qquad \Omega
^{2}=h-M^{2}+\frac{\sigma }{m_{\chi }^{2}},  \notag \\
\sigma \gamma &=&(1+\varepsilon )(1-2\varepsilon )(1-3\varepsilon )H^{4}+%
\dot{\varepsilon}\left( 1-6\varepsilon \right) H^{3}  \label{cd} \\
&&+\left[ m_{\chi }^{2}(1+\varepsilon )+\ddot{\varepsilon}+4\varepsilon 
\frac{k^{2}}{a^{2}}\right] H^{2}-H_{0}^{2}(H_{0}^{2}+m_{\chi }^{2}).  \notag
\end{eqnarray}

The constant $H_{0}$ is in principle arbitrary, but a remarkable choice, $%
H_{0}=m_{\phi }/2$, makes $\mathcal{L}_{\text{t}}^{(UV)}$ vanish in the de
Sitter limit. There, $U$ and $V$ decouple and 
\begin{equation}
(8\pi G)\frac{\mathcal{L}_{\text{t}}^{(U)}}{a^{3}\gamma }=\dot{U}^{2}-\frac{%
k^{2}}{a^{2}}U^{2},\qquad (8\pi GM^{4})\frac{\mathcal{L}_{\text{t}}^{(V)}}{%
a^{3}\gamma }=-\dot{V}^{2}+\left( \gamma m_{\chi }^{2}+\frac{k^{2}}{a^{2}}%
\right) V^{2}.  \label{ds}
\end{equation}

It is relatively straightforward to derive the power spectrum of the
fluctuations in this limit. The $V$ equation of motion is%
\begin{equation}
\ddot{V}+\frac{3}{2}m_{\phi }\dot{V}+\left( m_{\chi }^{2}+\frac{m_{\phi }^{2}%
}{2}+\frac{k^{2}}{a^{2}}\right) V=0.  \label{dseq}
\end{equation}%
As said, we need to solve it by means of the fakeon prescription and insert
the solution back into the action. Since (\ref{dseq}) is homogeneous and $U$%
-independent, the solution is just $V=0$.

Using $V=0$, formula (\ref{uU}) gives $u=U$, so we obtain a Mukhanov action%
\begin{equation*}
\mathcal{L}_{\text{t}}^{(U)}=\left( \frac{m_{\phi }^{2}+2m_{\chi }^{2}}{%
2m_{\chi }^{2}}\right) \mathcal{L}_{\text{t\hspace{0.01in}E}}^{(U)}
\end{equation*}
that coincides with the one of Einstein gravity with a scalar field, apart
from the overall factor. The $u$ two-point function in the de Sitter limit
is 
\begin{equation}
\langle u_{\mathbf{k}}(\tau )u_{\mathbf{k}^{\prime }}(\tau )\rangle =\frac{%
2m_{\chi }^{2}}{m_{\phi }^{2}+2m_{\chi }^{2}}\langle u_{\mathbf{k}}(\tau )u_{%
\mathbf{k}^{\prime }}(\tau )\rangle _{\text{E}},  \label{uu}
\end{equation}%
where%
\begin{equation*}
\langle u_{\mathbf{k}}(\tau )u_{\mathbf{k}^{\prime }}(\tau )\rangle _{\text{E%
}}=(2\pi )^{3}\delta ^{(3)}(\mathbf{k}+\mathbf{k}^{\prime })\frac{\pi
Gm_{\phi }^{2}}{2k^{3}}(1+k^{2}\tau ^{2}).
\end{equation*}%
Details on the derivation of (\ref{uu}) are given below. Formula (\ref{uu})
makes us already appreciate that the result depends on the mass $m_{\chi }$
of the fakeon in a nontrivial way.

\subsubsection*{Quasi de Sitter expansion}

Formulas (\ref{ab}) and (\ref{cd}) are exact, i.e., they do not assume $%
\varepsilon $ small. From now on, we work to the first order in $\varepsilon 
$, where we can use approximate formulas. Observe that (\ref{ab}) and (\ref%
{cd}) depend on $m_{\chi }$, $H$, $\varepsilon $ and $m_{\phi }$ (through $%
H_{0}=m_{\phi }/2$). However, the last three quantities are related by (\ref%
{ah}), so we can eliminate one of them. The price of this is that we
introduce terms proportional to $\sqrt{\varepsilon }$, which are unnecessary
at this level. It is possible to avoid it by switching to a slightly
different parametrization. Specifically, if we choose 
\begin{eqnarray}
\alpha &=&\frac{1}{M^{2}},\qquad \beta =M^{2}+m_{\chi }^{2}\gamma ,\qquad
f=3H-\frac{4\varepsilon H^{3}}{m_{\chi }^{2}\gamma },  \notag \\
h &=&M^{2}+m_{\chi }^{2}\gamma +\frac{k^{2}}{a^{2}}+\varepsilon \frac{%
H^{2}(m_{\chi }^{2}-4H^{2})}{m_{\chi }^{2}\gamma },\qquad \gamma =1+2\frac{%
H^{2}}{m_{\chi }^{2}},  \label{fis}
\end{eqnarray}%
and 
\begin{eqnarray}
\omega ^{2} &=&h-M^{2}-\frac{\sigma }{m_{\chi }^{2}}-m_{\chi }^{2}\gamma
,\qquad \Omega ^{2}=h-M^{2}+\frac{\sigma }{m_{\chi }^{2}},  \notag \\
\sigma \gamma &=&\varepsilon H^{2}\left( m_{\chi }^{2}-4H^{2}+\frac{4k^{2}}{%
\gamma a^{2}}\right) ,  \label{sigma}
\end{eqnarray}%
we find the Lagrangian 
\begin{equation}
(8\pi G)\frac{\mathcal{L}_{\text{t}}^{\prime }}{a^{3}\gamma }=\dot{U}%
^{2}-\omega ^{2}U^{2}+\frac{1}{M^{4}}\left( -\dot{V}^{2}+\Omega
^{2}V^{2}\right) +\frac{2\sigma }{m_{\chi }^{2}M^{2}}UV.  \label{ltg}
\end{equation}

The $V$ equation of motion is now%
\begin{equation}
\Sigma V=-\frac{\sigma M^{2}}{m_{\chi }^{2}}U,  \label{Veq}
\end{equation}%
where%
\begin{equation}
\Sigma \equiv \Sigma _{0}+\gamma m_{\chi }^{2},\qquad \Sigma _{0}\equiv 
\frac{\mathrm{d}^{2}}{\mathrm{d}t^{2}}+3H\frac{\mathrm{d}}{\mathrm{d}t}+%
\frac{k^{2}}{a^{2}}.  \label{sigm}
\end{equation}%
Anticipating that the solution for $V$ is of order $\varepsilon $, we have
dropped higher-order terms proportional to $\varepsilon V$, $\varepsilon 
\dot{V}$ from (\ref{Veq}). Let $\left. \Sigma ^{-1}\right\vert _{\text{f}}$
denote the fakeon Green function $G_{\text{f}}(t,t^{\prime })$, i.e., the
solution of $\Sigma G_{\text{f}}(t,t^{\prime })=\delta (t-t^{\prime })$
defined by the fakeon prescription (see section \ref{proj}). Then the
solution of (\ref{Veq}) can be written as 
\begin{equation}
\frac{V(t)}{M^{2}}=-\frac{1}{m_{\chi }^{2}}\int_{-\infty }^{+\infty }\mathrm{%
d}t^{\prime }\hspace{0.01in}G_{\text{f}}(t,t^{\prime })\sigma (t^{\prime
})U(t^{\prime })=-\frac{1}{m_{\chi }^{2}}\left. \Sigma ^{-1}\right\vert _{%
\text{f}}\sigma U.  \label{v}
\end{equation}

Inserting this expression into the Lagrangian (\ref{ltg}), we can see that
the nonlocal contribution due to $\mathcal{L}_{\text{t}}^{(UV)}$ is of order 
$\varepsilon ^{2}$, so we can drop it. The projected Lagrangian is 
\begin{equation}
(8\pi G)\frac{\mathcal{L}_{\text{t}}^{\text{prj}}}{a^{3}\gamma }=\dot{U}^{2}-%
\frac{\bar{k}^{2}}{a^{2}}U^{2},\qquad \bar{k}=k\left( 1-\frac{2\varepsilon
H^{2}}{m_{\chi }^{2}\gamma ^{2}}\right) .  \label{st}
\end{equation}

At this point, it is straightforward to work out the Mukhanov action.
Defining%
\begin{equation}
w=\frac{a\sqrt{\gamma }}{\sqrt{4\pi G}}U,\qquad \nu _{\text{t}}=\frac{3}{2}+%
\frac{\varepsilon }{\gamma },  \label{w}
\end{equation}%
and switching to the conformal time (\ref{tau}), the $w$ action to order $%
\varepsilon $ derived from (\ref{st}) reads%
\begin{equation}
S_{\text{t}}^{\text{prj}}=\frac{1}{2}\int \mathrm{d}\tau \left[ w^{\prime 2}-%
\bar{k}^{2}w^{2}+\frac{w^{2}}{\tau ^{2}}\left( \nu _{\text{t}}^{2}-\frac{1}{4%
}\right) \right] ,  \label{sred}
\end{equation}%
where the prime denotes the derivative with respect to $\tau $.

\subsubsection*{Power spectrum and spectral index}

Formula (\ref{sred}) tells us that the conjugate momentum of $w$ is $%
p=w^{\prime }$, so after turning $w$, $p$ into operators $\hat{w}$, $\hat{p}$%
, we impose the equal time quantization condition 
\begin{equation}
\lbrack \hat{p}_{\mathbf{k}}(\tau ),\hat{w}_{\mathbf{k}^{\prime }}(\tau
)]=-i\delta ^{(3)}\left( \mathbf{k}-\mathbf{k}^{\prime }\right) ,
\label{pqc}
\end{equation}%
where we have reinstated the subscripts $\mathbf{k}$. As usual, we write the
Fourier decomposition%
\begin{equation}
w_{\mathbf{k}}(\tau )=v_{\mathbf{k}}(\tau )\hat{a}_{\mathbf{k}}+v_{-\mathbf{k%
}}^{\ast }(\tau )\hat{a}_{-\mathbf{k}}^{\dagger },\qquad \lbrack \hat{a}_{%
\mathbf{k}},\hat{a}_{\mathbf{k}^{\prime }}^{\dagger }]=(2\pi )^{3}\delta
^{(3)}(\mathbf{k}-\mathbf{k}^{\prime }),  \label{aac}
\end{equation}%
where $\hat{a}_{\mathbf{k}}^{\dagger }$ and $\hat{a}_{\mathbf{k}}$ are
creation and annihilation operators.

The limit $k/(aH)\rightarrow \infty $ of (\ref{sred}) allows us to define
the Bunch-Davies vacuum state $|0\rangle $. From formula (\ref{sred}), we
see that the only difference with respect to the result obtained in the de
Sitter limit is a rescaling of $k$. Thus, we require%
\begin{equation}
v_{\mathbf{k}}\rightarrow \frac{\mathrm{e}^{-i\bar{k}\tau }}{\sqrt{2\bar{k}}}%
\qquad \text{for }\frac{k}{aH}\rightarrow \infty .  \label{vacu}
\end{equation}%
Using the condition\ (\ref{vacu}), we can work out the modes $v_{\mathbf{k}}$
and obtain%
\begin{equation}
U_{\mathbf{k}}=\pi H(1-\varepsilon )|\tau |^{3/2}\sqrt{\frac{G}{\gamma }}%
\left[ \mathrm{e}^{i\pi (2\nu _{\text{t}}+1)/4}H_{\nu _{\text{t}}}^{(1)}(|%
\bar{k}\tau |)\hat{a}_{\mathbf{k}}+\mathrm{e}^{-i\pi (2\nu _{\text{t}%
}+1)/4}H_{\nu _{\text{t}}}^{(2)}(|\bar{k}\tau |)\hat{a}_{-\mathbf{k}%
}^{\dagger }\right] ,  \label{Uk}
\end{equation}%
having used the third formula of (\ref{ah}), where $H_{\nu _{\text{t}%
}}^{(1,2)}$ are the Hankel functions. For the purpose of computing the power
spectrum, we need to work out the leading behavior in the superhorizon limit 
$|k\tau |\rightarrow 0$.\ There we have%
\begin{equation}
U_{\mathbf{k}}=(1-\varepsilon )\left( \frac{|\bar{k}\tau |}{2}\right)
^{(3-2\nu _{\text{t}})/2}\frac{H\Gamma (\nu _{\text{t}})}{\bar{k}^{3/2}}%
\sqrt{\frac{8G}{\gamma }}\left[ \mathrm{e}^{i\pi (2\nu _{\text{t}}-1)/4}\hat{%
a}_{\mathbf{k}}+\mathrm{e}^{-i\pi (2\nu _{\text{t}}-1)/4}\hat{a}_{-\mathbf{k}%
}^{\dagger }\right] .  \label{uk}
\end{equation}

The redefinitions (\ref{uU}) tell us that to compute the $u$ two-point
function we also need the fakeon $V_{\mathbf{k}}$, which is given by formula
(\ref{v}). While the general discussion of the fakeon Green function is left
to section \ref{proj}, here we can quickly get to the result we need as
follows. In the superhorizon limit $|k\tau |\rightarrow 0$ we can ignore the
term proportional to $k^{2}/a^{2}$ in the expression (\ref{sigma}) of $%
\sigma $. Once we do this, we can commute $\sigma $ and $\left. \Sigma
^{-1}\right\vert _{\text{f}}$ in (\ref{v}), because the commutator gives
corrections of higher orders in $\varepsilon $. Moreover, recalling that $%
\Sigma _{0}U_{\mathbf{k}}=\mathcal{O}(\varepsilon )$, because $U_{\mathbf{k}%
} $ solves the Mukhanov equation of the projected Lagrangian $\mathcal{L}_{%
\text{t}}^{\text{prj}}$ of formula (\ref{st}), $\left. \Sigma
^{-1}\right\vert _{\text{f}}$ just multiplies $U_{\mathbf{k}}$ by $1/(\gamma
m_{\chi }^{2})$. Collecting these facts, we have, in the superhorizon limit
and discarding higher orders,%
\begin{equation*}
\frac{V_{\mathbf{k}}}{M^{2}}=-\frac{1}{m_{\chi }^{2}}\left. \Sigma
^{-1}\right\vert _{\text{f}}\sigma U_{\mathbf{k}}=-\frac{\sigma }{m_{\chi
}^{2}}\left. \Sigma ^{-1}\right\vert _{\text{f}}U_{\mathbf{k}}=-\frac{\sigma 
}{m_{\chi }^{2}}\left. \frac{1}{\Sigma _{0}+\gamma m_{\chi }^{2}}\right\vert
_{\text{f}}U_{\mathbf{k}}=-\frac{\sigma }{m_{\chi }^{2}}\frac{1}{\gamma
m_{\chi }^{2}}U_{\mathbf{k}},
\end{equation*}%
that is to say,%
\begin{equation}
\frac{V_{\mathbf{k}}}{M^{2}}=-\frac{\varepsilon H^{2}}{m_{\chi }^{4}\gamma
^{2}}(m_{\chi }^{2}-4H^{2})U_{\mathbf{k}}.  \label{Vk}
\end{equation}

The power spectrum $\mathcal{P}_{u}$ of each graviton polarization is
defined by%
\begin{equation}
\langle u_{\mathbf{k}}(\tau )u_{\mathbf{k}^{\prime }}(\tau )\rangle =(2\pi
)^{3}\delta ^{(3)}(\mathbf{k}+\mathbf{k}^{\prime })\frac{2\pi ^{2}}{k^{3}}%
\mathcal{P}_{u}.  \label{deltau}
\end{equation}%
The two-point function can be evaluated in the superhorizon limit from (\ref%
{uU}), (\ref{Uk}) and (\ref{Vk}). We find%
\begin{equation*}
\mathcal{P}_{u}=\frac{GH^{2}}{\pi \gamma }\left( 1-\frac{2\varepsilon }{%
\gamma }+\frac{2\varepsilon }{\gamma }\psi _{0}(3/2)\right) \left( \frac{%
k|\tau |}{2}\right) ^{-2\varepsilon /\gamma },
\end{equation*}%
where $\psi _{0}$ is the digamma function.

The power spectrum of the tensor fluctuations, matched with the usual
conventions, is $\mathcal{P}_{T}=16\mathcal{P}_{u}$. Replacing $|\tau |$ by $%
1/k_{\ast }$, where $k_{\ast }$ is a reference scale, it is common to write 
\begin{equation}
\ln \mathcal{P}_{T}(k)=\ln A_{T}+n_{T}\ln \frac{k}{k_{\ast }},  \label{lnpr}
\end{equation}%
where $A_{T}$ and $n_{T}$ are called amplitude and spectral index (or tilt),
respectively. We find%
\begin{eqnarray}
A_{T} &=&\frac{8Gm_{\chi }^{2}m_{\phi }^{2}}{\pi (m_{\phi }^{2}+2m_{\chi
}^{2})}\left( 1-\frac{2\sqrt{3\varepsilon }m_{\chi }^{2}}{m_{\phi
}^{2}+2m_{\chi }^{2}}-\frac{\varepsilon m_{\chi }^{2}(2m_{\chi
}^{2}+37m_{\phi }^{2})}{6(m_{\phi }^{2}+2m_{\chi }^{2})^{2}}-n_{T}(2-\gamma
_{E}-\ln 2)\right) ,\qquad  \notag \\
n_{T} &=&3-2\nu _{\text{t}}=-\frac{4\varepsilon m_{\chi }^{2}}{m_{\phi
}^{2}+2m_{\chi }^{2}},  \label{spec}
\end{eqnarray}%
where $\gamma _{E}$ is the Euler-Mascheroni constant and we have used the
first formula of (\ref{ah}) to eliminate $H$.

\subsection{Scalar fluctuations}

\label{scalarocm}

Now we study the scalar fluctuations in the inflaton framework. We work in
the comoving gauge, where the $\phi $ fluctuation $\delta \phi $ is set to
zero and the metric reads 
\begin{equation}
g_{\mu \nu }=\text{diag}(1,-a^{2},-a^{2},-a^{2})+2\text{diag}(\Phi
,a^{2}\Psi ,a^{2}\Psi ,a^{2}\Psi )-\delta _{\mu }^{0}\delta _{\nu
}^{i}\partial _{i}B-\delta _{\mu }^{i}\delta _{\nu }^{0}\partial _{i}B.
\label{mets}
\end{equation}

After Fourier transforming the space coordinates, (\ref{sqgeq}) gives the
quadratic Lagrangian 
\begin{eqnarray}
(8\pi G)\frac{\mathcal{L}_{\text{s}}}{a^{3}} &=&-3(\dot{\Psi}+H\Phi
)^{2}+4\pi G\dot{\phi}^{2}\Phi ^{2}+\frac{k^{2}}{a^{2}}\left[ 2B(\dot{\Psi}%
+H\Phi )+\Psi (\Psi -2\Phi )\right]  \notag \\
&&-\frac{k^{4}}{3a^{4}m_{\chi }^{2}}\left[ (\dot{B}+\Phi +\Psi
)^{2}-2BH(\Phi +\Psi )-4\pi G\dot{\phi}^{2}B^{2}\right] .  \label{lsMD}
\end{eqnarray}%
As before, $\Psi ^{2}$ stands for $\Psi _{-\mathbf{k}}\Psi _{\mathbf{k}}$, $%
\dot{\Psi}^{2}$ for $\dot{\Psi}_{-\mathbf{k}}\dot{\Psi}_{\mathbf{k}}$, and
so on.

Since $\Phi $ appears algebraically, we eliminate it by means of its own
field equation. We remain with a Lagrangian that depends only on $B$ and $%
\Psi $. The field redefinitions 
\begin{equation}
\Psi =\frac{U}{\sqrt{\varepsilon }},\qquad \qquad B=\frac{a^{2}}{k^{2}}V+%
\frac{3U}{\sqrt{\varepsilon }H(3-\varepsilon )},  \label{uU2}
\end{equation}%
allow us to decompose $\mathcal{L}_{\text{s}}$ as 
\begin{equation}
\mathcal{L}_{\text{s}}=\mathcal{L}_{\text{s}}^{(U)}+\mathcal{L}_{\text{s}%
}^{(V)}+\mathcal{L}_{\text{s}}^{(UV)},  \label{lps}
\end{equation}%
where $\mathcal{L}_{\text{s}}^{(UV)}$ is the sum of a term proportional to $%
UV$ plus one proportional to $\dot{U}V$. In addition, $\mathcal{L}_{\text{s}%
}^{(UV)}$ vanishes in the de Sitter limit.

We do not give the full expression of $\mathcal{L}_{\text{s}}$ here, but
stress that after the redefinition (\ref{uU2}) it admits a series expansion
in powers of $k$ and $\sqrt{\varepsilon }$. In particular,%
\begin{equation*}
\lim_{\varepsilon \rightarrow 0,k\rightarrow 0}(8\pi G)\frac{\mathcal{L}_{%
\text{s}}}{a^{3}}=\dot{U}^{2}-\frac{1}{3m_{\chi }^{2}}\left( \dot{V}%
^{2}-m_{\chi }^{2}\gamma V^{2}\right) ,
\end{equation*}%
where $\gamma $ is defined in (\ref{fis}). As before, $V$ is the fakeon and $%
U$ is the physical excitation.

\subsubsection*{Quasi de Sitter expansion}

In the de Sitter limit $\varepsilon \rightarrow 0$, we find%
\begin{eqnarray}
(8\pi G)\frac{\mathcal{L}_{\text{s}}^{(U)}}{a^{3}} &=&\dot{U}^{2}-\frac{k^{2}%
}{a^{2}}U^{2},\qquad \mathcal{L}_{\text{s}}^{(UV)}=0,  \notag \\
(24\pi Gm_{\chi }^{2}\upsilon )\frac{\mathcal{L}_{\text{s}}^{(V)}}{a^{3}}
&=&-\dot{V}^{2}+\left[ 18H^{4}+3H^{2}m_{\chi }^{2}(3-\hat{k}^{2}+2\hat{k}%
^{4})+\hat{k}^{4}m_{\chi }^{4}(1+\hat{k}^{2})\right] \frac{V^{2}}{9\upsilon
H^{2}},\qquad  \label{lv}
\end{eqnarray}%
where 
\begin{equation*}
\upsilon =1+\frac{\hat{k}^{4}m_{\chi }^{2}}{9H^{2}},\qquad \hat{k}=\frac{k}{%
m_{\chi }a}.
\end{equation*}%
Note that $\upsilon $ and the coefficient of $V^{2}$ in (\ref{lv}) are
positive definite.

Again, we see that the fakeon $V$ decouples. Its own equation of motion sets
it to zero, so the Lagrangian of $U$ coincides with the usual Mukhanov
expression, normalization included. This means that the power spectrum of
the scalar fluctuations coincides with the one of Einstein gravity in this
limit.

To order $\eta \sim \sqrt{\varepsilon }$, we find%
\begin{eqnarray}
(8\pi G)\frac{\mathcal{L}_{\text{s}}^{(U)}}{a^{3}} &=&\dot{U}^{2}-\frac{k^{2}%
}{a^{2}}U^{2}+2\sqrt{3\varepsilon }H^{2}U^{2},  \notag \\
(8\pi G)\frac{\mathcal{L}_{\text{s}}^{(UV)}}{a^{3}} &=&\frac{2\sqrt{%
\varepsilon }V}{9\upsilon }\left[ (2\hat{k}^{2}-3)\dot{U}+\frac{\hat{k}%
^{2}m_{\chi }^{2}}{9\upsilon H^{3}}\left( 9H^{2}-12\hat{k}^{2}H^{2}+8\hat{k}%
^{4}H^{2}+\hat{k}^{4}m_{\chi }^{2}\right) U\right] .\qquad  \label{lve}
\end{eqnarray}%
Since the $V$ equation of motion implies $V=\mathcal{O}(\sqrt{\varepsilon })$%
, $\mathcal{L}_{\text{s}}^{(V)}$ remains the one of formula (\ref{lv}) to
the order we are considering. Moreover, after integrating $V$ out, the
projected $U$ Lagrangian is just $\mathcal{L}_{\text{s}}^{(U)}$, since the $%
V $-dependent corrections are $\mathcal{O}(\varepsilon )$.

From $\mathcal{L}_{\text{s}}^{(U)}$, we can derive the Mukhanov action by
following the steps from (\ref{w}) to (\ref{uk}), with the replacements $%
\bar{k}\rightarrow k$, $\gamma \rightarrow 1$ and $\nu _{\text{t}%
}\rightarrow \nu _{\text{s}}$, where 
\begin{equation}
\nu _{\text{s}}=\frac{3}{2}+2\sqrt{\frac{\varepsilon }{3}}.  \label{nus}
\end{equation}%
Recalling that in the comoving gauge the curvature perturbation $\mathcal{R}$
coincides with $\Psi $, we can derive the power spectrum $\mathcal{P}_{%
\mathcal{R}}$, defined by%
\begin{equation}
\langle \mathcal{R}_{\mathbf{k}}(\tau )\mathcal{R}_{\mathbf{k}^{\prime
}}(\tau )\rangle =(2\pi )^{3}\delta ^{(3)}(\mathbf{k}+\mathbf{k}^{\prime })%
\frac{2\pi ^{2}}{k^{3}}\mathcal{P}_{\mathcal{R}}.  \label{psi2}
\end{equation}%
Inserting the solution for $U$ into the left formula of (\ref{uU2}), we find%
\begin{equation}
\ln \mathcal{P}_{\mathcal{R}}(k)=\ln A_{\mathcal{R}}+(n_{\mathcal{R}}-1)\ln 
\frac{k}{k_{\ast }},  \label{lnprr}
\end{equation}%
where the amplitude $A_{\mathcal{R}}$ and the spectral index $n_{\mathcal{R}%
}-1$ are 
\begin{eqnarray}
A_{\mathcal{R}} &=&\frac{Gm_{\phi }^{2}}{4\pi \varepsilon }\left( 1-\sqrt{%
3\varepsilon }-(n_{\mathcal{R}}-1)(2-\gamma _{E}-\ln 2)\right) ,
\label{depsi} \\
n_{\mathcal{R}}-1 &=&3-2\nu _{\text{s}}=-4\sqrt{\frac{\varepsilon }{3}},
\label{ns}
\end{eqnarray}%
respectively. We see that the mass $m_{\chi }$ of the fakeon does not affect
the result to the order we are considering.

Finally, from (\ref{spec}) we derive the tensor-to-scalar ratio 
\begin{equation}
r=\frac{A_{T}}{A_{\mathcal{R}}}=\frac{32\varepsilon m_{\chi }^{2}}{m_{\phi
}^{2}+2m_{\chi }^{2}}\left( 1+\frac{\sqrt{3\varepsilon }m_{\phi }^{2}}{%
m_{\phi }^{2}+2m_{\chi }^{2}}+(n_{\mathcal{R}}-1-n_{T})(2-\gamma _{E}-\ln
2)\right) .  \label{rati}
\end{equation}

\section{The fakeon projection}

\label{proj}\setcounter{equation}{0}

In this section we discuss the fakeon projection, starting from the tensor
fluctuations. The Lagrangian $\mathcal{L}_{\text{t}}^{(V)}$ of formula (\ref%
{ltg}) leads to the $V$ equation of motion (\ref{Veq}). The fakeon Green
function $G_{\text{f}}(t,t^{\prime })$ is the solution of $\Sigma G_{\text{f}%
}(t,t^{\prime })=\delta (t-t^{\prime })$, defined by the fakeon
prescription, where $\Sigma $ is given in formula (\ref{sigm}). For the
purposes of this paper, it is sufficient to invert $\Sigma $ in the de
Sitter limit $a(t)=\mathrm{e}^{Ht}$, where $H$ is treated as a constant. We
keep $H$ generic to make the discussion easily adaptable to the geometric
framework. We will use the information that $H$ is $m_{\phi }/2$ in the de
Sitter limit (in the inflaton framework) only later.

It is convenient to switch to a symmetric operator by noting that%
\begin{equation}
\Sigma a^{-3}=a^{-3/2}\left( \frac{\mathrm{d}^{2}}{\mathrm{d}t^{2}}+m_{\chi
}^{2}-\frac{H^{2}}{4}+\frac{k^{2}}{a^{2}}\right) a^{-3/2}.  \label{sh}
\end{equation}%
We want to prove that the fakeon solution $\hat{G}_{\text{f}}(t,t^{\prime })$
of 
\begin{equation}
\left( \frac{\mathrm{d}^{2}}{\mathrm{d}t^{2}}+m_{\chi }^{2}-\frac{H^{2}}{4}+%
\frac{k^{2}}{a^{2}}\right) \hat{G}_{\text{f}}(t,t^{\prime })=\delta
(t-t^{\prime })  \label{mas}
\end{equation}%
is 
\begin{equation}
\hat{G}_{\text{f}}(t,t^{\prime })=\frac{i\pi \mathrm{sgn}(t-t^{\prime })}{%
4H\sinh \left( n_{\chi }\pi \right) }\left[ J_{in_{\chi }}(\check{k}%
)J_{-in_{\chi }}(\check{k}^{\prime })-J_{in_{\chi }}(\check{k}^{\prime
})J_{-in_{\chi }}(\check{k})\right] ,  \label{feG}
\end{equation}%
where $\mathrm{sgn}(t)$ is the sign function, $J_{n}$ denotes the Bessel
function of the first kind and 
\begin{equation}
n_{\chi }=\sqrt{\frac{m_{\chi }^{2}}{H^{2}}-\frac{1}{4}},\qquad \check{k}=%
\frac{k}{a(t)H},\qquad \check{k}^{\prime }=\frac{k}{a(t^{\prime })H}.
\label{kappa}
\end{equation}

In principle, we could add solutions of the homogeneous equation, which are
the functions $J_{\pm in_{\chi }}(\check{k})$, multiplied by constants. The
job of the projection is to determine those constants uniquely. Because it
comes from quantum field theory, the fakeon projection is known
perturbatively around flat space, in four-momentum space. However, a notion
of four-momentum is not immediately available in curved space.

Fortunately, there are three limits where $\hat{G}_{\text{f}}$ is known,
which are $k/(aH)\rightarrow \infty $, $k/(aH)=0$ and $a=$ constant. The
limit $k/(aH)\rightarrow \infty $ gives the flat-space case once we switch
to conformal time. The limit $k/(aH)\rightarrow 0$ gives the flat-space case
if we keep the cosmological time. The case $a=$ constant is precisely flat
space, but is not relevant here, since we are interested in the de Sitter
background. Hence, necessary conditions are that the solution (\ref{feG})
reduces to the known expressions \cite{causalityQG,FLRW} in both cases $%
k/(aH)\rightarrow \infty $ and $k/(aH)=0$. Any of these two conditions is
also sufficient. The other condition can be seen as a consistency check.

Switching to conformal time $\tau =-1/(aH)$, equation (\ref{mas}) can be
written as 
\begin{equation*}
\left( \frac{\mathrm{d}^{2}}{\mathrm{d}\tau ^{2}}+k^{2}+\frac{m_{\chi }^{2}}{%
\tau ^{2}H^{2}}\right) \left( H\sqrt{\tau \tau ^{\prime }}\hat{G}_{\text{f}%
}\right) =\delta (\tau -\tau ^{\prime }).
\end{equation*}%
For $k|\tau |$ large we obtain%
\begin{equation}
\left( \frac{\mathrm{d}^{2}}{\mathrm{d}\tau ^{2}}+k^{2}\right) \left( H\sqrt{%
\tau \tau ^{\prime }}\hat{G}_{\text{f}}\right) \simeq \delta (\tau -\tau
^{\prime }).  \label{klarge1}
\end{equation}%
Solving it by means of the arithmetic average of the retarded and advanced
potentials, we find \cite{causalityQG,FLRW} 
\begin{equation}
\hat{G}_{\text{f}}\simeq \frac{1}{2Hk\sqrt{\tau \tau ^{\prime }}}\sin \left(
k|\tau -\tau ^{\prime }|\right) .  \label{feGtau}
\end{equation}%
It is easy to check that (\ref{feG}) does satisfy (\ref{feGtau}) when $%
k|\tau |,k|\tau ^{\prime }|\gg 1$.

As said, the most general solution of (\ref{mas}) is equal to (\ref{feG})
plus solutions of the homogeneous equation, multiplied by constant
coefficients $c_{1}$ and $c_{2}$. Now we know that those coefficients must
vanish, to match (\ref{feGtau}) for $k|\tau |,k|\tau ^{\prime }|$ large.
This proves that (\ref{feG}) is the correct fakeon Green function.

A consistency check is given by the limit $k\rightarrow 0$. There, (\ref{mas}%
) turns into an equation similar to (\ref{klarge1}), provided we keep the
cosmological time $t$ instead of switching to $\tau $. Consequently, the
solution (\ref{feG}) must tend to \cite{causalityQG,FLRW} 
\begin{equation}
\frac{1}{2Hn_{\chi }}\sin \left( Hn_{\chi }|t-t^{\prime }|\right) .
\label{fe}
\end{equation}%
It is easy to check that this is indeed the $k\rightarrow 0$ limit of (\ref%
{feG}).

From (\ref{sh}) we derive the fakeon Green function%
\begin{equation}
G_{\text{f}}(t,t^{\prime })=\frac{i\pi \mathrm{sgn}(t-t^{\prime })\mathrm{e}%
^{-3H(t-t^{\prime })/2}}{4H\sinh \left( n_{\chi }\pi \right) }\left[
J_{in_{\chi }}(\check{k})J_{-in_{\chi }}(\check{k}^{\prime })-J_{in_{\chi }}(%
\check{k}^{\prime })J_{-in_{\chi }}(\check{k})\right] .  \label{feGG}
\end{equation}

\subsection{Consistency condition}

We have determined the fakeon Green function in curved space by referring to
two situations where the problem becomes a flat-space one, which are $%
k/(aH)\rightarrow \infty $ and $k/(aH)\rightarrow 0$. As mentioned in the
introduction, purely virtual particles are subject to a consistency
(no-tachyon) condition in flat-space, i.e., their squared mass should be
positive. Formula (\ref{feGtau}) shows that this requirement is always
satisfied for $k/(aH)\rightarrow \infty $, while formula (\ref{fe}) shows
that it is satisfied for $k/(aH)\rightarrow 0$ if $n_{\chi }$ is real.
Recalling that $H$ is $m_{\phi }/2$ in the inflaton\ framework, the
condition reads%
\begin{equation}
m_{\chi }>\frac{m_{\phi }}{4},  \label{notachyon}
\end{equation}%
which is a lower bound on the mass of the fakeon with respect to the mass of
the inflaton. When (\ref{notachyon}) holds, the oscillating behavior of (\ref%
{fe}) suppresses the contributions with 
\begin{equation*}
|t-t^{\prime }|\gg \frac{1}{Hn_{\chi }}=\frac{4}{\sqrt{16m_{\chi
}^{2}-m_{\phi }^{2}}}.
\end{equation*}

One may wonder if it is meaningful to impose a condition stronger than (\ref%
{notachyon}), for example require that the time-dependent squared mass be
positive for all values of $k/(aH)$. To discuss this issue, let us consider
the Lagrangian that gives the fakeon Green function of formula (\ref{mas}),
which is%
\begin{equation}
\hat{L}=-\frac{1}{2}\left( \frac{\mathrm{d}\hat{V}}{\mathrm{d}t}\right) ^{2}+%
\frac{m(t)^{2}}{2}\hat{V}^{2},\qquad m(t)^{2}=m_{\chi }^{2}-\frac{H^{2}}{4}+%
\frac{k^{2}}{a^{2}}.  \label{lV}
\end{equation}%
A redefinition $t=h(t^{\prime })$, $\hat{V}(t)= f(t^{\prime })\tilde{V}%
(t^{\prime })$, with $\mathrm{d}h/\mathrm{d}t^{\prime }=f^{2}$, leaves the
kinetic term invariant, but changes the squared mass. Specifically, the
transformed Lagrangian reads 
\begin{equation*}
\tilde{L}=-\frac{1}{2}\left( \frac{\mathrm{d}\tilde{V}}{\mathrm{d}t^{\prime }%
}\right) ^{2}+\frac{M(t^{\prime })^{2}}{2}\tilde{V}^{2},
\end{equation*}%
where%
\begin{equation}
M^{2}=f^{4}m^{2}-f\frac{\mathrm{d}^{2}}{\mathrm{d}t^{\prime 2}}\left( \frac{1%
}{f}\right) =f^{4}m^{2}+f^{3}\frac{\mathrm{d}^{2}f}{\mathrm{d}t^{2}}.
\label{masstr}
\end{equation}%
This transformation law shows that the signs of $m(t)^{2}$ and $M(t^{\prime
})^{2}$ do not have a reparametri\-zation-independent meaning, in general,
so a squared mass that becomes negative in some time interval is not
necessarily a sign of a lack of consistency.

In passing, it is easy to verify that if the masses are independent of time,
then the condition of positive square mass is independent of the
parametrization. Indeed, if $m(t)^{2}$ is $t$-independent and positive, the
most general reparametrization $f\left( t^{\prime }\right) $ that leaves $%
M(t^{\prime })^{2}$ $t^{\prime }$-independent has%
\begin{equation*}
f(t^{\prime }(t))^{2}=\sqrt{\rho ^{2}+\frac{M^{2}}{m^{2}}}+\rho \cos
(2mt+\theta ),
\end{equation*}%
where $\rho $ and $\theta $ are arbitrary real constants of integration.
Since $f^{2}$ must be real and identically positive, $M^{2}$ must also be
positive.

Summarizing, a necessary condition for the fakeon projection in the
inflationary scenario is that the fakeon squared mass be positive in the
superhorizon limit: 
\begin{equation}
\left. m(t)^{2}\right\vert _{k/(aH)\rightarrow 0}>0.  \label{consistency}
\end{equation}

This condition also leads to (\ref{notachyon}) in the case of the scalar
fluctuations. Indeed, consider the Lagrangian $\mathcal{L}_{\text{s}}^{(V)}$
given in formula (\ref{lv}). Making the change of variables 
\begin{equation}
V(t)=\frac{\sqrt{\upsilon }}{a^{3/2}}W(t),  \label{fw}
\end{equation}%
the $W$ equation of motion takes the form%
\begin{equation}
\ddot{W}+m(t)^{2}W=\mathcal{O}(\sqrt{\varepsilon }),  \label{ww}
\end{equation}%
for some involved rational function $m(t)^{2}$ of $H^{2}/m_{\chi }^{2}$ and $%
k^{2}/(aH)^{2}$, equal to $(4m_{\chi }^{2}-H^{2})/4$ in the superhorizon
limit. Thus, (\ref{consistency}) gives again the bound (\ref{notachyon}).

As we show in section \ref{vector}, the vector fluctuations give the same
bound. The same bound is also found in the geometric framework. It is
conceivable that, if (\ref{notachyon}) were violated, the theory would
predict a rather different large-scale structure of the universe, or a
different scenario would have to be envisaged to produce the present
situation.

The stronger requirement that $m(t)^{2}$ be positive for every $k$ makes
sense if we believe that the cosmological time plays a special role. Then we
still find the bound (\ref{notachyon}) for the tensor fluctuations, while a
stronger bound is obtained in the case of the scalar fluctuations. Studying
the coefficient $m(t)^{2}$ of $W$ in (\ref{ww}) numerically, we find that it
is positive for all values of $k^{2}/(aH)^{2}$ if%
\begin{equation}
m_{\chi }\geqslant 0.312\hspace{0.01in}m_{\phi }.  \label{notas}
\end{equation}%
As soon as $m_{\chi }\lesssim 0.312\hspace{0.01in}m_{\phi }$, there exists a
finite $k$ domain where $m(t)^{2}$ has negative values. When $m_{\chi }$
satisfies (\ref{notachyon}) but not (\ref{notas}), there is a time interval $%
\Delta t\sim \ln (k/m_{\phi })/m_{\phi }$, comparable with the duration of
inflation, where the fakeon Green function is \textquotedblleft
tachyonic\textquotedblright\ and its nonlocal contribution is no longer
negligible.

In the rest of the paper, we take (\ref{notachyon}) as the consistency
condition for the fakeon projection in inflationary cosmology, because it is
universal and reparametrization independent. Yet, the issues just mentioned
suggest that there is a chance that it might be conservative. The formulas
of the power spectra do not depend on it, but (\ref{notas}) narrows the
window of allowed values of the tensor-to-scalar ratio $r$ a little bit more
than (\ref{notachyon}) (see section \ref{predictions}).

\section[Geometric framework]{Geometric framework ($R+R^{2}+C^{2}$)}

\label{geometric}\setcounter{equation}{0}

In this section we study the geometric framework, which is sometimes known
in the literature as Jordan frame. The higher-derivative equations of the
background metric, derived from (\ref{SQG}) with the FLRW\ ansatz, can be
written in the simple form%
\begin{equation}
\frac{\dot{\varepsilon}}{H}=-3\varepsilon \left( 1-\frac{\varepsilon }{2}%
\right) +\frac{m_{\phi }^{2}}{2H^{2}},  \label{bac}
\end{equation}%
where $\varepsilon $ is again $-\dot{H}/H^{2}$. It is worth to stress that $%
\varepsilon $, $H$, $a$ and the cosmological time $t$ are different from
those of the inflaton framework, although we denote them by means of the
same symbols. The match between the two frameworks is worked out in detail
in Appendix \ref{matching}.

The quasi de Sitter approximation of (\ref{bac}) requires $\varepsilon \sim
m_{\phi }^{2}/(6H^{2})\ll 1$, so $H$ is no longer related to $m_{\phi }$ in
the de Sitter limit, where actually $m_{\phi }\ll H$. As far as the mass $%
m_{\chi }$ is concerned, it can be either of order $H$ or of order $m_{\phi
} $. This means that we have two types of quasi de Sitter expansions,
depending on whether $m_{\chi }\sim H$ or $m_{\chi }\sim m_{\phi }$. We
study the scalar and tensor fluctuations in both.

The two possibilities can also be understood as follows. The de Sitter
metric is not an exact solution of the field equations of the theory $%
R+R^{2}+C^{2}$. It is an exact solution in two cases: ($i$) when we ignore
both $R$ and $C^{2}$; and ($ii$) when we ignore just $R$. In other words,
the term $R^{2}$ is leading with respect to the term $R$, while the term $%
C^{2}$ can either be of order $R$ or of order $R^{2}$ (as far as the
fluctuations are concerned). The first case is studied by expanding in
powers of $\varepsilon $ with $\xi =H^{2}/m_{\chi }^{2}$ fixed. The second
case is studied by expanding in powers of $\varepsilon $ with $\zeta \equiv
m_{\chi }^{2}/m_{\phi }^{2}$ fixed.

The relation between $m_{\phi }$, $H$ and $\varepsilon $ is%
\begin{equation}
\frac{m_{\phi }^{2}}{H^{2}}=\varepsilon \left( 6+\varepsilon -\frac{2}{3}%
\varepsilon ^{2}\right) +\mathcal{O}(\varepsilon ^{4}).  \label{appro}
\end{equation}%
It can be found by writing down the most general expansion for $m_{\phi
}^{2}/H^{2}$ in powers of $\varepsilon $, differentiating it and applying (%
\ref{bac}) to determine the coefficients. If needed, (\ref{appro}) can be
extended to arbitrarily high orders (an asymptotic series being obtained).

\subsection[Tensor fluctuations]{$m_{\chi }\sim H$: tensor fluctuations}

\label{tensorHD}

We start from the tensor fluctuations. Parametrizing the metric as in (\ref%
{met}), the quadratic Lagrangian obtained from (\ref{SQG}) is 
\begin{equation}
(8\pi G)\frac{\mathcal{L}_{\text{t}}}{a^{3}}=\dot{u}^{2}\left( 1+\frac{%
2\Upsilon }{m_{\phi }^{2}}+\frac{\Upsilon }{m_{\chi }^{2}}+\frac{2k^{2}}{%
a^{2}m_{\chi }^{2}}\right) -\frac{k^{2}}{a^{2}}\left( 1+\frac{2\Upsilon }{%
m_{\phi }^{2}}+\frac{k^{2}}{a^{2}m_{\chi }^{2}}\right) u^{2}-\frac{\ddot{u}%
^{2}}{m_{\chi }^{2}},  \label{ltHD}
\end{equation}%
plus an identical contribution for $v$, where%
\begin{equation}
\Upsilon \equiv 2H^{2}+\dot{H}.  \label{ups}
\end{equation}

Expanding around the de Sitter background with $\xi =H^{2}/m_{\chi }^{2}$
fixed, the first nonvanishing contribution to the spectral index $n_{T}$
turns out to be $\mathcal{O}(\varepsilon ^{2})$. For this reason, we work
out the predictions to the second order in $\varepsilon $ included.
Expanding the Lagrangian (\ref{ltHD}), we find%
\begin{eqnarray}
(8\pi G)\frac{\mathcal{L}_{\text{t}}}{a^{3}} &=&\dot{U}^{2}\left( 1+\frac{%
5\varepsilon }{6}+\frac{2\varepsilon ^{2}}{9}+3\varepsilon \xi +\frac{15}{2}%
\varepsilon ^{2}\xi +\frac{3k^{2}\varepsilon }{a^{2}m_{\chi }^{2}}\right)
-\left( 3+2\varepsilon +9\varepsilon \xi \right) \varepsilon H^{2}U^{2} 
\notag \\
&&-\frac{k^{2}U^{2}}{a^{2}}\left( 1+\frac{5\varepsilon }{6}+\frac{%
2\varepsilon ^{2}}{9}+3\varepsilon ^{2}\xi +\frac{3k^{2}\varepsilon }{%
2a^{2}m_{\chi }^{2}}\right) -\frac{3\varepsilon \ddot{U}^{2}}{2m_{\chi }^{2}}%
,  \label{UHD}
\end{eqnarray}%
where%
\begin{equation*}
U=\sqrt{\frac{2}{3\varepsilon }}u.
\end{equation*}

The important point of (\ref{UHD}) is that the unique higher-derivative term 
$\ddot{U}^{2}$ is multiplied by $\varepsilon $, so the fakeon projection can
be handled iteratively. The change of variables%
\begin{eqnarray}
U &=&E\left( 1-\frac{5\varepsilon }{12}+\frac{43\varepsilon ^{2}}{288}%
\right) -\frac{3}{2}\varepsilon \xi E\left( 1-\frac{7\varepsilon }{4}\right)
\notag \\
&&+\frac{27}{8}\varepsilon ^{2}\xi ^{2}E-\frac{9\varepsilon \xi }{4H}\dot{E}%
+\varepsilon \mathcal{O}(k^{2}|\tau |^{2},\varepsilon \dot{E},\ddot{E}%
,\cdots )  \label{ch}
\end{eqnarray}%
allows us to cast the Lagrangian in the form%
\begin{equation*}
(8\pi G)\frac{\mathcal{L}_{\text{t}}}{a^{3}}=\dot{E}^{2}-\frac{k^{2}}{a^{2}}%
E^{2}-3\varepsilon \left( 1-\varepsilon \right) H^{2}E^{2}.
\end{equation*}

Since $\dot{E}\simeq \varepsilon E$ for $|k\tau |$ small [as in (\ref{uk})],
the corrections $\varepsilon \mathcal{O}(k^{2}|\tau |^{2},\varepsilon \dot{E}%
,\ddot{E},\cdots )$ of (\ref{ch}) are either $\mathcal{O}(\varepsilon
^{5/2}) $ or give subleading contributions in the superhorizon limit $%
k\left\vert \tau \right\vert \ll 1$. This means that we do not need\ to
specify them for our purposes.

At this point, it is sufficient to upgrade the steps from formula (\ref{w})
to formula (\ref{uk}) to the appropriate order, with the substitutions $%
U\rightarrow E$, $\bar{k}\rightarrow k$, $\gamma \rightarrow 1$. We find 
\begin{equation}
\nu _{\text{t}}=\frac{3}{2}+3\varepsilon ^{2}.  \label{nutHD}
\end{equation}%
The power spectrum of the tensor fluctuations is $\mathcal{P}_{T}=16\mathcal{%
P}_{u}$, with $\mathcal{P}_{u}$ defined by (\ref{deltau}). Using the
definition (\ref{lnpr}), the amplitude and the spectral index are%
\begin{eqnarray}
A_{T} &=&\frac{24GH^{2}}{\pi }\varepsilon \left[ 1-\frac{17\varepsilon }{6}%
-3\varepsilon \xi -\frac{31}{36}\varepsilon ^{2}+17\varepsilon ^{2}\xi
+9\varepsilon ^{2}\xi ^{2}-n_{T}(2-\gamma _{E}-\ln 2)\right] ,
\label{specHD} \\
n_{T} &=&\frac{\mathrm{d}\ln \mathcal{P}_{T}(k)}{\mathrm{d}\ln k}=3-2\nu _{%
\text{t}}=-6\varepsilon ^{2}.  \label{tiltHD}
\end{eqnarray}

\subsection[Scalar fluctuations]{$m_{\chi }\sim H$: scalar fluctuations}

\label{scalarHD}

Now we discuss the scalar fluctuations in the geometric framework by
expanding in powers of $\varepsilon $ to the next-to-leading order with $\xi
=H^{2}/m_{\chi }^{2}$ fixed. We switch directly from (\ref{SQG}) to the
action (\ref{SQGmix}) (with $S_{\mathfrak{m}}\rightarrow 0$), to remove the
higher derivatives without changing the metric that couples to matter. We
isolate the background value of $\varphi $ from its fluctuation $\Omega $ by
writing 
\begin{equation}
\varphi =-6\Upsilon +\Omega ,  \label{fihat}
\end{equation}%
where $\Upsilon $ is defined in (\ref{ups}). The gauge invariant curvature
perturbation $\mathcal{R}$ is%
\begin{equation}
\mathcal{R}=\Psi -\frac{H}{6\dot{\Upsilon}}\Omega .  \label{R}
\end{equation}

We work in the spatially-flat gauge, where $\Omega $ is an independent field
and $\Psi $ is set to zero. This means that the metric is 
\begin{equation}
g_{\mu \nu }=\text{diag}(1+2\Phi ,-a^{2},-a^{2},-a^{2})-\delta _{\mu
}^{0}\delta _{\nu }^{i}\partial _{i}B-\delta _{\mu }^{i}\delta _{\nu
}^{0}\partial _{i}B.  \label{meta}
\end{equation}%
After Fourier transforming the space coordinates, the quadratic Lagrangian
reads 
\begin{eqnarray}
(8\pi G)\frac{\mathcal{L}_{\text{s}}}{a^{3}} &=&\frac{1}{m_{\phi }^{2}}\left[
(\Upsilon -H^{2})\Phi \Omega -H\Phi \dot{\Omega}-3\Upsilon ^{2}\Phi ^{2}-%
\frac{\Omega ^{2}}{12}\right] +\frac{k^{2}}{a^{2}}HB\Phi  \notag \\
&&-\frac{k^{2}}{3Hm_{\phi }^{2}a^{2}}\left[ H\Omega (\Phi +\dot{B}%
)+2H^{2}B\Omega -3\Upsilon B\Phi (\Upsilon +2H^{2})\right]  \notag \\
&&-\frac{k^{4}}{3a^{4}m_{\chi }^{2}}(\Phi +\dot{B}-BH)^{2}.  \label{lascal}
\end{eqnarray}

The field $\Phi $ appears in (\ref{lascal}) as a Lagrange multiplier, so we
integrate it out by solving its own field equation and inserting the
solution back into the action. So doing, we obtain a two-derivative
quadratic Lagrangian for $B$ and $\Omega $, which we then expand around the
de Sitter background by means of (\ref{appro}). Making the field
redefinitions%
\begin{equation}
\Omega =12\sqrt{2\varepsilon }H^{2}U,\qquad B=\frac{3a^{2}}{k^{2}}\sqrt{%
\frac{\varepsilon }{2}}V,  \label{Ome}
\end{equation}%
we obtain an action that is regular for $\varepsilon ,k\rightarrow 0$. Its $%
\varepsilon =0$ limit is%
\begin{equation*}
(8\pi G)\lim_{\varepsilon \rightarrow 0}\frac{\mathcal{L}_{\text{s}}}{a^{3}}=%
\dot{U}^{2}+V^{2}-\frac{k^{2}}{a^{2}}\left( U^{2}+\frac{2UV}{3H}-\frac{%
k^{2}U^{2}}{9a^{2}H^{2}}\right) .
\end{equation*}%
We note that at this level $V$ appears algebraically and can be integrated
out. This means that the fakeon projection can be handled iteratively in $%
\varepsilon $.

After integrating $V$ out, every $m_{\chi }$ dependence disappears to the
first order in $\varepsilon $. In particular, if we define 
\begin{equation}
U=\left( 1-\frac{5}{12}\varepsilon \right) E,  \label{Omu}
\end{equation}%
the action becomes%
\begin{equation}
(8\pi G\mathcal{)}\frac{\mathcal{L}_{\text{s}}}{a^{3}}=\dot{E}^{2}-\frac{%
k^{2}}{a^{2}}E^{2}+3\varepsilon H^{2}E^{2}.  \label{lauaHD}
\end{equation}

The redefinition (\ref{w}) with $U\rightarrow E$, $\gamma \rightarrow 1$ and 
$\nu _{\text{t}}\rightarrow \nu _{\text{s}}$, where 
\begin{equation}
\nu _{\text{s}}=\frac{3}{2}+2\varepsilon ,  \label{nusHD}
\end{equation}%
gives the action (\ref{sred}) with $\bar{k}\rightarrow k$. Inserting the
solution for $E$ into (\ref{Omu}), (\ref{Ome}) and then (\ref{R}), and using
the definition (\ref{lnprr}), we find in the superhorizon limit, 
\begin{eqnarray}
A_{\mathcal{R}} &=&\frac{GH^{2}}{2\pi \varepsilon }\left( 1-\frac{17}{6}%
\varepsilon -(n_{\mathcal{R}}-1)(2-\gamma _{E}-\ln 2)\right) ,  \label{des}
\\
n_{\mathcal{R}}-1 &=&\frac{\mathrm{d}\ln \mathcal{P}_{\mathcal{R}}}{\mathrm{d%
}\ln k}=3-2\nu _{\text{s}}=-4\varepsilon .  \label{ndHD}
\end{eqnarray}

Together with (\ref{specHD}), formula (\ref{des}) gives the tensor-to-scalar
ratio 
\begin{equation}
r=\frac{A_{T}}{A_{\mathcal{R}}}=48\varepsilon ^{2}\left( 1-3\varepsilon \xi
-4\varepsilon (2-\gamma _{E}-\ln 2)\right) ,  \label{rsHD}
\end{equation}%
to the next-to-leading order in $\varepsilon $. More explicitly, we get,
after inverting (\ref{appro}),%
\begin{equation}
r=\frac{96m_{\chi }^{2}\varepsilon ^{2}}{m_{\phi }^{2}+2m_{\chi }^{2}}\left(
1-4\varepsilon (2-\gamma _{E}-\ln 2)\right) .  \label{final}
\end{equation}

So far, we have assumed $\varepsilon $ small and $\xi $ arbitrary. However,
we see from (\ref{specHD}) and (\ref{rsHD}) that higher orders of $%
\varepsilon $ carry higher powers of $\xi $. To write (\ref{final}) we have
used%
\begin{equation*}
\frac{1}{1+3\varepsilon \xi }=1-3\varepsilon \xi +9\varepsilon ^{2}\xi ^{2}+%
\mathcal{O}(\varepsilon ^{3}\xi ^{3}).
\end{equation*}%
Conservatively, formula (\ref{final}) is reliable as long as $3\varepsilon
\xi \simeq m_{\phi }^{2}/(2m_{\chi }^{2})$ is reasonably smaller than one.
However, we may argue that the overall factor in front of (\ref{final}) is
exact. In the next two subsections we show that it is indeed so.

\subsection[Tensor fluctuations 2]{$m_{\chi }\sim m_{\phi }$: tensor
fluctuations}

\label{tensorHD2}

Now we study the tensor fluctuations in the geometric framework with $\zeta
=m_{\chi }^{2}/m_{\phi }^{2}$ fixed. The metric is still parametrized as (%
\ref{met}) and the quadratic Lagrangian obtained from (\ref{SQG}) is (\ref%
{ltHD}), plus an identical contribution for $v$. After replacing $m_{\chi
}^{2}$ with $m_{\phi }^{2}\zeta $, we use (\ref{appro}) to eliminate $%
m_{\phi }^{2}$ and then expand in $\varepsilon $. We work out the leading
and next-to-leading orders in $\varepsilon $.

As in subsection \ref{tensor}, we eliminate the higher derivatives of (\ref%
{ltHD}) by considering the extended Lagrangian $\mathcal{L}_{\text{t}%
}^{\prime }=\mathcal{L}_{\text{t}}+\Delta \mathcal{L}_{\text{t}}$, where $%
\Delta \mathcal{L}_{\text{t}}$ is defined in (\ref{dlt}). If we perform the
redefinitions%
\begin{equation*}
u=\sqrt{\frac{3\varepsilon \zeta }{1+2\zeta }}\left( 1-\frac{5\varepsilon }{%
12}\right) (U+V),\qquad S=2\sqrt{3\varepsilon \zeta (1+2\zeta )}H^{2}\left(
1-\frac{5\varepsilon }{12}\right) (U-\varepsilon V),
\end{equation*}%
and choose%
\begin{equation}
f=3H,\qquad h=2(1+2\zeta )H^{2}+\frac{k^{2}}{a^{2}},  \label{values}
\end{equation}%
we obtain 
\begin{eqnarray}
(8\pi G)\frac{\mathcal{L}_{\text{t}}^{(U)}}{a^{3}} &=&\dot{U}^{2}-\frac{k^{2}%
}{a^{2}}U^{2}\left( 1-\frac{2\varepsilon }{1+2\zeta }\right) -3\varepsilon
H^{2}U^{2},  \notag \\
(8\pi G)\frac{\mathcal{L}_{\text{t}}^{(V)}}{a^{3}} &=&-\dot{V}%
^{2}+hV^{2},\qquad (8\pi G)\frac{\mathcal{L}_{\text{t}}^{(UV)}}{a^{3}}%
=4\varepsilon V\left( H\dot{U}+\frac{k^{2}}{a^{2}}\frac{U}{1+2\zeta }\right)
.  \notag
\end{eqnarray}

As usual, we have just written the $\varepsilon \rightarrow 0$ limit of $%
\mathcal{L}_{\text{t}}^{(V)}$, since the fakeon projection implies $V=%
\mathcal{O}(\varepsilon )$. This means that, to the order of approximation
we are considering, we can drop both $\mathcal{L}_{\text{t}}^{(V)}$ and $%
\mathcal{L}_{\text{t}}^{(UV)}$, so the projected Lagrangian is just $%
\mathcal{L}_{\text{t}}^{(U)}$.

Switching to conformal time and defining%
\begin{equation}
w=\frac{aU}{\sqrt{4\pi G}},\qquad \nu _{\text{t}}=\frac{3}{2},\qquad \bar{k}%
=k\left( 1-\frac{\varepsilon }{1+2\zeta }\right) ,  \label{wHD}
\end{equation}%
the Mukhanov action is (\ref{sred}). The fakeon Green function $G_{\text{f}%
}(t,t^{\prime })$ can be discussed as in subsection \ref{proj}, with the
replacements 
\begin{equation}
m_{\chi }^{2}\rightarrow 4\zeta H^{2},\qquad n_{\chi }\rightarrow \sqrt{%
\frac{4m_{\chi }^{2}}{m_{\phi }^{2}}-\frac{1}{4}}.  \label{nchiHD}
\end{equation}%
and the solution is still (\ref{feGG}). The consistency condition (\ref%
{consistency}) gives again (\ref{notachyon}). Aside from the changes (\ref%
{nchiHD}), everything works as before and we find, from the $V$ field
equation of $\mathcal{L}_{\text{t}}^{\prime }$, 
\begin{equation}
V(t)=-2\varepsilon \int_{-\infty }^{+\infty }\mathrm{d}t^{\prime }\hspace{%
0.01in}G_{\text{f}}(t,t^{\prime })\left[ H\dot{U}(t^{\prime })+\frac{k^{2}}{%
a(t^{\prime })^{2}}\frac{U(t^{\prime })}{1+2\zeta }\right] .  \label{VtHD2}
\end{equation}

The fakeon average can be worked out with the procedure of subsection \ref%
{tensor}. Recalling that the terms in the square bracket of (\ref{VtHD2})
are subleading or of higher orders in $\varepsilon $, we obtain that $V$
does not contribute in the superhorizon limit $|k\tau |\ll 1$.

Inverting (\ref{appro}) to restore the $m_{\phi }^{2}$ dependence of the
overall factor, the power spectrum $\mathcal{P}_{T}=16\mathcal{P}_{u}$ of
the tensor fluctuations gives the amplitude%
\begin{equation}
A_{T}=\frac{8G}{\pi }\frac{m_{\chi }^{2}m_{\phi }^{2}}{m_{\phi
}^{2}+2m_{\chi }^{2}}\left( 1-\frac{6\varepsilon m_{\chi }^{2}}{m_{\phi
}^{2}+2m_{\chi }^{2}}\right) ,  \label{specHD2}
\end{equation}%
while the spectral index $n_{T}$ is $\mathcal{O}(\varepsilon ^{2})$.

\subsection[Scalar fluctuations 2]{$m_{\chi }\sim m_{\phi }$: scalar
fluctuations}

\label{scalarHD2}

Now we study the scalar fluctuations in the geometric framework with $\zeta
=m_{\chi }^{2}/m_{\phi }^{2}$ fixed. We replace $m_{\chi }^{2}$ with $%
m_{\phi }^{2}\zeta $, use (\ref{appro}) to eliminate $m_{\phi }^{2}$ and
then expand in powers of $\varepsilon $. We work to the next-to-leading
order in $\varepsilon $.

We eliminate the higher derivatives of (\ref{SQG}) by means of (\ref{SQGmix}%
). The metric is still parametrized as (\ref{meta}) in the spatially-flat
gauge $\Psi =0$. The curvature perturbation is (\ref{R}) and the $\varphi $
fluctuation $\Omega $ is defined by (\ref{fihat}). The quadratic Lagrangian
obtained from (\ref{SQGmix}) is (\ref{lascal}). Defining%
\begin{equation*}
\Omega =12\sqrt{2\varepsilon }H^{2}U\left( 1-\frac{5\varepsilon }{12}\right)
,\qquad B=\frac{V}{\sqrt{2}k^{2}H}\sqrt{\varepsilon \left( k^{4}+36\zeta
a^{4}H^{4}\right) }+\sqrt{\frac{\varepsilon }{2}}\frac{U}{H}\left( 1+\frac{%
7\varepsilon }{12}\right) ,
\end{equation*}%
and expanding to the next-to-leading order in $\varepsilon $, we obtain the
decomposition (\ref{lps}) with 
\begin{eqnarray}
(8\pi G)\frac{\mathcal{L}_{\text{s}}^{(U)}}{a^{3}} &=&\dot{U}^{2}-\frac{k^{2}%
}{a^{2}}U^{2}+3\varepsilon H^{2}U^{2},  \notag \\
(8\pi G)\frac{\mathcal{L}_{\text{s}}^{(V)}}{a^{3}} &=&-\dot{V}^{2}+\left[
2(1+2\zeta )+\frac{k^{2}}{a^{2}H^{2}}+g_{1}\right] H^{2}V^{2},\qquad
\label{ls2} \\
(8\pi G)\frac{\mathcal{L}_{\text{s}}^{(UV)}}{a^{3}} &=&\varepsilon V\left(
g_{2}H\dot{U}+\frac{k^{2}}{a^{2}}g_{3}U\right) ,  \notag
\end{eqnarray}%
where $g_{i}$, $i=1,2,3$, are regular functions of $k/(aH)$ and $\zeta $,
which tend to finite values in both limits $k/(aH)\rightarrow 0,\infty $.
Moreover, $g_{1}$ tends to zero for $k/(aH)\rightarrow 0$ and $g_{3}$ tends
to zero for $k/(aH)\rightarrow \infty $.

The expression of $\mathcal{L}_{\text{s}}^{(V)}$ of (\ref{ls2}) is to the
leading order, which is sufficient for our present purposes. The discussion
about the fakeon Green function proceeds as before. It is easy to check that
the consistency condition (\ref{consistency}) coincides with (\ref{notachyon}%
). Clearly, the fakeon projection implies $V=\mathcal{O}(\varepsilon )$.
This means that the projected Lagrangian is just $\mathcal{L}_{\text{s}%
}^{(U)}$ to the order of approximation we are considering, i.e., we can drop
both $\mathcal{L}_{\text{s}}^{(V)}$ and $\mathcal{L}_{\text{s}}^{(UV)}$.

The Mukhanov action is (\ref{sred}) with $\gamma \rightarrow 1$, $\bar{k}%
\rightarrow k$ and $\nu _{\text{t}}\rightarrow \nu _{\text{s}%
}=(3/2)+2\varepsilon $. The power spectrum $\mathcal{P}_{\mathcal{R}}$ gives
the amplitude%
\begin{equation}
A_{\mathcal{R}}=\frac{m_{\phi }^{2}G}{12\pi \varepsilon ^{2}}\left(
1-3\varepsilon -(n_{\mathcal{R}}-1)(2-\gamma _{E}-\ln 2)\right)
\label{specHDs}
\end{equation}%
and the spectral index $n_{\mathcal{R}}-1=-4\varepsilon $. Again, the
dependence on $m_{\chi }$ drops out. Combining this result with (\ref%
{specHD2}), the tensor-to-scalar ratio is%
\begin{equation}
r=\frac{96m_{\chi }^{2}\varepsilon ^{2}}{m_{\phi }^{2}+2m_{\chi }^{2}}\left(
1+\frac{3\varepsilon m_{\phi }^{2}}{m_{\phi }^{2}+2m_{\chi }^{2}}%
-4\varepsilon (2-\gamma _{E}-\ln 2)\right) ,  \label{final2}
\end{equation}%
which agrees with (\ref{final}) for $\zeta $ large.

\section{Vector fluctuations}

\label{vector}\setcounter{equation}{0}

In this section we study the vector fluctuations and show that they are set
to zero by the fakeon projection at the quadratic level. For definiteness,
we work in the geometric framework, but equivalent results are obtained in
the inflaton framework.

We parametrize the metric as 
\begin{equation*}
g_{\mu \nu }=\text{diag}(1,-a^{2},-a^{2},-a^{2})-\delta _{\mu }^{0}\delta
_{\nu }^{i}B_{i}-\delta _{\mu }^{i}\delta _{\nu }^{0}B_{i}-\delta _{\mu
}^{i}\delta _{\nu }^{j}(\partial _{i}E_{j}+\partial _{j}E_{i}),
\end{equation*}%
where $\partial ^{i}B_{i}=0$ and $\partial ^{i}E_{i}=0$. A gauge invariant
quantity is 
\begin{equation}
B_{i}-\dot{E}_{i}  \label{ginv}
\end{equation}%
We choose a gauge where $E_{i}=0$ and rewrite the metric as 
\begin{equation}
g_{\mu \nu }=\text{diag}(1,-a^{2},-a^{2},-a^{2})-\delta _{\mu }^{0}\delta
_{\nu }^{1}C-\delta _{\mu }^{0}\delta _{\nu }^{2}D-\delta _{\mu }^{1}\delta
_{\nu }^{0}C-\delta _{\mu }^{2}\delta _{\nu }^{0}D,  \label{vectormet}
\end{equation}%
where $C=C(t,z)$ and $D=D(t,z)$ are the independent vector modes. After
Fourier transforming the space coordinates, the quadratic Lagrangian $%
\mathcal{L}_{v}$ obtained from (\ref{SQG}) is given by 
\begin{equation}
(32\pi Gm_{\chi }^{2}a)\frac{\mathcal{L}_{v}}{k^{2}}=-\dot{C}^{2}+\left[
m_{\chi }^{2}+\left( 4\zeta +\varepsilon -2\varepsilon \zeta \right) H^{2}+%
\frac{k^{2}}{a^{2}}\right] C^{2}  \label{lvec}
\end{equation}%
plus an identical contribution for $D$, where $\zeta =m_{\chi }^{2}/m_{\phi
}^{2}$. As before, $C^{2}$ stands for $C_{-\mathbf{k}}C_{\mathbf{k}}$, $\dot{%
C}^{2}$ for $\dot{C}_{-\mathbf{k}}\dot{C}_{\mathbf{k}}$, and so on. After
the redefinition%
\begin{equation*}
\mathcal{V}=\frac{kC}{2m_{\chi }a^{1/2}},
\end{equation*}%
the Lagrangian turns into 
\begin{equation}
(8\pi G)\mathcal{L}_{v}=-\dot{\mathcal{V}}^{2}+\left[ m_{\chi }^{2}+(16\zeta
-1+2\varepsilon -8\varepsilon \zeta )\frac{H^{2}}{4}+\frac{k^{2}}{a^{2}}%
\right] \mathcal{V}^{2}.  \label{vecact}
\end{equation}%
The kinetic term has the wrong sign, so $\mathcal{V}$ needs to be quantized
as a fakeon. Since $\mathcal{V}$ does not couple to any other field at this
level, the fakeon projection sets it to zero. Therefore, the vector modes do
not contribute to the two-point functions. Note that these conclusions hold
without expanding around the de Sitter background.

The consistency condition (\ref{consistency}) is studied by requiring that
the coefficient of $\mathcal{V}^{2}$ in (\ref{vecact}) be positive in the
superhorizon/de Sitter limit, which gives again (\ref{notachyon}).

\section{Summary of predictions and connection with observations}

\label{predictions}\setcounter{equation}{0}

In this section we summarize the predictions and make contact with
observations. We express the results in terms of the number of $e$-foldings,
which is defined by 
\begin{equation}
N=\int_{t_{i}}^{t_{f}}H(t^{\prime })\mathrm{d}t^{\prime },  \label{n}
\end{equation}%
where $t_{i}$ is the time when $\varepsilon (t_{i})=\varepsilon $ and $t_{f}$
is when inflation ends, $\varepsilon (t_{f})=1$. It is convenient to work in
the geometric framework, where we can use (\ref{bac}) and (\ref{appro}).
Then we translate the formulas to the inflaton framework by means of the map
of appendix \ref{matching}. Expressing every quantity as a function of $%
\varepsilon $, (\ref{n}) gives 
\begin{equation}
N=\int_{\varepsilon }^{1}\frac{H(t^{\prime }(\varepsilon ^{\prime }))}{\dot{%
\varepsilon}(t^{\prime }(\varepsilon ^{\prime }))}\mathrm{d}\varepsilon
^{\prime }=\int_{\varepsilon }^{1}\frac{\mathrm{d}\varepsilon ^{\prime }}{%
2\varepsilon ^{\prime 2}}\left[ 1+\frac{\varepsilon ^{\prime }}{6}+\mathcal{O%
}(\varepsilon ^{\prime 2})\right] =\frac{1}{2\varepsilon }-\frac{1}{12}\ln
\varepsilon +\text{ }\mathcal{O}(\varepsilon ^{0}).  \label{N}
\end{equation}%
The $\mathcal{O}(\varepsilon ^{0})$ corrections are not very meaningful,
because they depend on the upper bound of integration and $\varepsilon
(t_{f})=1$ is just a conventional choice. To the leading order, we can take%
\begin{equation*}
N\simeq \frac{1}{2\varepsilon },
\end{equation*}%
in the geometric framework. Note that in the inflaton framework we have
instead $N\simeq \sqrt{3}/(2\sqrt{\varepsilon })$, as can be shown using (%
\ref{epsa}). Once expressed in terms of $N$, the predictions obtained in the
two frameworks agree (see appendix \ref{matching}). Collecting the results
of formulas (\ref{spec}), (\ref{depsi})-(\ref{ns}), (\ref{specHD2}) and (\ref%
{specHDs}), we obtain, to the leading order, 
\begin{equation}
\renewcommand{\arraystretch}{2.5}{%
\begin{tabular}{|c|c|c|c|c|}
\hline
{\large $A_{\mathcal{R}}$} & {\large $A_{T}$} & {\large $r$} & {\large $~n_{%
\mathcal{R}}-1~ $} & {\large $n_{T}$} \\[.5ex] \hline
{\Large $~\frac{m_{\phi }^{2}N^{2}}{3\pi M_{\text{Pl}}^{2}}~ $} & {\Large $~%
\frac{8m_{\chi }^{2}m_{\phi }^{2}}{\pi (m_{\phi }^{2}+2m_{\chi }^{2})M_{%
\text{Pl}}^{2}}~ $} & {\Large ~$\frac{24m_{\chi }^{2}}{N^{2}(m_{\phi
}^{2}+2m_{\chi }^{2})} $~} & {\Large ~$-\frac{2}{N} $~} & {\Large ~$-\frac{%
3m_{\chi }^{2}}{N^{2}(m_{\phi }^{2}+2m_{\chi }^{2})} $~} \\[1.5ex] \hline
\end{tabular}%
}  \label{table}
\end{equation}
The formula of $n_{T}$ comes from (\ref{spec}), since in this particular
case the inflaton framework is more powerful than the geometric framework.

We see that the predictions for $A_{\mathcal{R}}$ and $n_{\mathcal{R}}-1$
coincide with the ones of the $R+R^{2}$ model. Instead, the predictions for $%
A_{T}$, $r$ and $n_{T}$ are smaller by a factor $2m_{\chi }^{2}/(m_{\phi
}^{2}+2m_{\chi }^{2})$. Note that (\ref{table}) implies the relation%
\begin{equation}
r\simeq -8n_{T},  \label{preda}
\end{equation}%
which is known to hold in single-field slow-roll models independently of the
scalar potential $V(\phi )$ \cite{peters}. It is a nontrivial fact that it
does not depend on $m_{\chi }$, besides $N$ and $m_{\phi }$.

\begin{figure}[t]
\begin{center}
\includegraphics[width=10truecm]{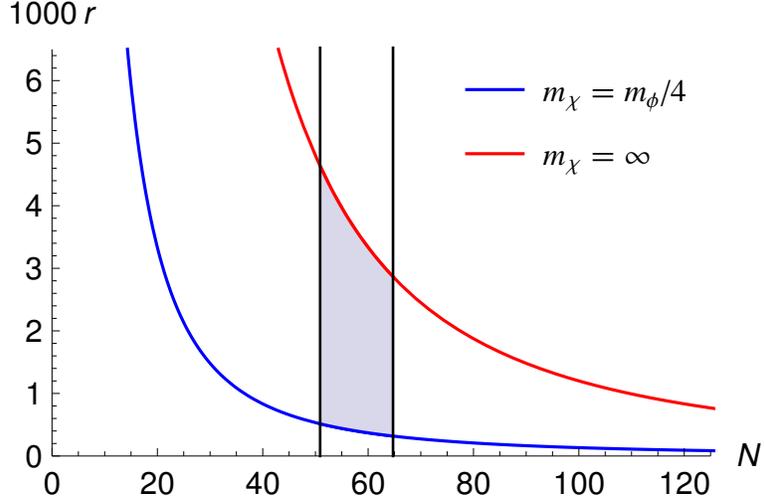}
\end{center}
\caption{Allowed values of the tensor-to-scalar ratio $r$}
\label{plot}
\end{figure}

The bound (\ref{notachyon}) on $m_{\chi }$ is also a prediction of the
theory, required by the consistency of the fakeon projection with
inflationary cosmology. Because of it, the tensor-to-scalar ratio $r$ and
the spectral index $n_{T}$ are predicted within less than one order of
magnitude. Precisely, 
\begin{equation}
\frac{1}{9}\lesssim \frac{N^{2}}{12}r\simeq -\frac{2N^{2}}{3}n_{T}\lesssim 1.
\label{prediction}
\end{equation}%
For example, for $N=60$ we have 
\begin{equation*}
0.4\lesssim 1000r\lesssim 3,\qquad -0.4\lesssim 1000n_{T}\lesssim -0.05.
\end{equation*}%
The allowed values of $r$ are shown in fig. \ref{plot}, where the vertical
lines denote the minimum and maximum values of $N$ in the range $n_{\mathcal{%
R}}=0.9649\pm 0.0042$ at 68\% CL \cite{Planck18}. The windows (\ref%
{prediction}) are compatible with the data available at present, which give $%
r<0.1$ \cite{Planck18}.

The results of this paper also provide corrections to the amplitudes, which
can be used to estimate the theoretical errors. From (\ref{N}) we find%
\begin{equation*}
\varepsilon =\frac{1}{2N}\left( 1+\frac{\ln N}{12N}\right) +\mathcal{O}%
\left( \frac{1}{N^{2}}\right) ,
\end{equation*}%
so we obtain 
\begin{eqnarray*}
A_{\mathcal{R}} &=&\frac{GN^{2}m_{\phi }^{2}}{3\pi }\left( 1-\frac{\ln N}{6N}%
+\mathcal{O}\left( \frac{1}{N}\right) \right) . \\
A_{T} &=&\frac{8G}{\pi }\frac{m_{\chi }^{2}m_{\phi }^{2}}{m_{\phi
}^{2}+2m_{\chi }^{2}}\left( 1-\frac{3m_{\chi }^{2}}{N(m_{\phi }^{2}+2m_{\chi
}^{2})}\left( 1+\frac{\ln N}{12N}\right) +\mathcal{O}\left( \frac{1}{N^{2}}%
\right) \right) .
\end{eqnarray*}%
With $N=60$, the first correction to $A_{T}$ is between 0.3\% ($m_{\chi
}=m_{\phi }/4$) and 2.5\% ($m_{\chi }\rightarrow \infty $). Although $A_{%
\mathcal{R}}$ and $n_{\mathcal{R}}-1$ do not depend on $m_{\chi }$ in our
approximation, they will at higher orders.

\section{Conclusions}

\label{conclusions}\setcounter{equation}{0}

We have worked out the predictions of quantum gravity with fakeons on
inflationary cosmology. By expanding around the de Sitter background the
amplitudes and spectral indices of the scalar and tensor fluctuations have
been calculated to the next-to-leading orders, comparing different
frameworks, which lead to matching results. The physical content of the
theory is exhausted by the two power spectra. The vector degrees of freedom,
as well as the other scalar and tensor ones, are handled by means of the
fakeon prescription and projected away. The methodologies we have developed
to deal with this operation appear to be generalizable to higher orders.

The local, renormalizable, unitary, perturbative quantum field theory of
gravity considered in this paper depends only on four parameters: the
cosmological constant, Newton's constant, $m_{\phi }$ and $m_{\chi }$. The
values of the cosmological constant and Newton's constant are known. It will
be possible to derive the values of $m_{\phi }$ and $m_{\chi }$ from $n_{%
\mathcal{R}}$ and $r$ once new cosmological data will be available \cite%
{CMBStage4}. At that point, the theory will be uniquely determined and all
other predictions (tensor tilt, running of the spectral indices, and so on)
will be stringent tests of its validity.

The consistency of the approach puts a lower bound on the mass $m_{\chi }$
of the fakeon with respect to the mass $m_{\phi }$ of the scalar field. The
tensor-to-scalar ratio $r$ is determined within less than an order of
magnitude. Moreover, the relation $r=-8n_{T}$ is not affected by $m_{\chi }$
within our approximation. A separate analysis is required to study the case
where the consistency bound on $m_{\chi }$ is violated and work out the
consequences of the violation on the physics of the primordial universe.
Finally, the investigation of this paper and the results we have obtained
shed light on the problem of understanding purely virtual particles in
curved space.

\vskip 1truecm \noindent {\large \textbf{Acknowledgments}}

\vskip .5truecm

We are grateful to Denis Comelli and Gianfranco Cordella for helpful
discussions. E.B. is supported by the NSF Grant No. PHY-1806428 and
acknowledges the support of the ID 61466 Grant from the John Templeton
Foundation (JTF), as part of the QISS project. M.P. is supported by the
Estonian Research Council grants PRG803 and MOBTT86 and by the EU through
the European Regional Development Fund CoE program TK133 \textquotedblleft
The Dark Side of the Universe\textquotedblright . M.P. is also grateful to
Fondazione Angelo Della Riccia for financial support during the early stage
of this work.

\vskip 1truecm

\noindent{\textbf{\huge Appendices}} \renewcommand{\thesection}{%
\Alph{section}} \renewcommand{\theequation}{\thesection.\arabic{equation}} %
\setcounter{section}{0}

\section{Map relating the inflaton\ framework to the geometric framework}

\label{matching} \setcounter{equation}{0}

In this appendix we derive some expansions used in the paper and show that
the results of the inflaton and geometric frameworks agree with each other.
For definiteness, the quantities with bars ($\bar{a}$, $\bar{t}$, $\bar{H}$, 
$\bar{\varepsilon}$, $\bar{\eta}$, etc.) refer to the inflaton framework,
while the quantities without bars ($a$, $t$, $H$, $\varepsilon $, $\eta $,
etc.) refer to the geometric framework.

We start by determining $-aH\tau $ in the geometric formalism. Writing the
most general expansion%
\begin{equation}
-aH\tau =1+a_{1}\varepsilon +a_{2}\varepsilon ^{2}+a_{3}\varepsilon
^{3}+\cdots ,  \label{ansa}
\end{equation}%
the numerical coefficients $a_{i}$ are calculated by differentiating (\ref%
{ansa}) and then using the definition (\ref{tau}), the equation (\ref{bac})
and (\ref{ansa}) again. This procedure gives an equality of two power
series. Matching the coefficients recursively, we obtain $a_{i}$ for every $%
i $. To the lowest orders, the result is 
\begin{equation}
-aH\tau =1+\varepsilon +3\varepsilon ^{2}+\frac{44}{3}\varepsilon ^{3}+%
\mathcal{O}(\varepsilon ^{4}).  \label{ahHD}
\end{equation}%
If we continue to arbitrary orders, we find an asymptotic series.

The action (\ref{sqgeq}) is obtained from (\ref{SQG}) by means of the
conformal transformation (\ref{weyl}). If we want to map the
parametrizations (\ref{met}) and (\ref{mets}) of the background metrics into
each other, we need to combine that transformation with a time redefinition $%
\bar{t}(t)$, so that 
\begin{equation}
\frac{\text{\textrm{d}}\bar{t}}{\text{\textrm{d}}t}=\frac{\bar{a}}{a},\qquad 
\text{\textrm{d}}\bar{s}^{2}=\bar{g}_{\mu \nu }\mathrm{d}\bar{x}^{\mu }%
\mathrm{d}\bar{x}^{\nu }=\mathcal{W}\text{\textrm{d}}s^{2}=\mathcal{W}g_{\mu
\nu }\mathrm{d}x^{\mu }\mathrm{d}x^{\nu },\qquad \mathcal{W}\equiv 1-\frac{%
\varphi }{3m_{\phi }^{2}}.  \label{weylmetric}
\end{equation}

We split the conformal factor $\mathcal{W}$ into the sum of its background
part $\mathcal{W}_{0}$ and the fluctuation $\delta \mathcal{W}$. Using (\ref%
{fihat}) and the second equation of (\ref{weyl}), it is easy to find 
\begin{equation}
\mathcal{W}_{0}=1+\frac{2\Upsilon }{m_{\phi }^{2}},\qquad \delta \mathcal{W}%
=-\frac{\Omega }{3m_{\phi }^{2}}=-\hat{\kappa}\mathcal{W}_{0}\delta \phi .
\label{relationsW}
\end{equation}%
The transformations of the background quantities are 
\begin{equation}
\frac{\text{\textrm{d}}\bar{t}}{\text{\textrm{d}}t}=\frac{\bar{a}}{a}=\sqrt{%
\mathcal{W}_{0}},\qquad \bar{H}=\frac{1}{\sqrt{\mathcal{W}_{0}}}\left( H+%
\frac{\dot{\mathcal{W}}_{0}}{2\mathcal{W}_{0}}\right) ,  \label{mdsubs}
\end{equation}%
plus those of $\varepsilon $ and $\eta $, which follow directly from their
definitions. Using (\ref{bac}), (\ref{ups}) and (\ref{appro}), we find, to
the lowest orders, 
\begin{eqnarray}
\bar{H} &=&\frac{m_{\phi }}{2}\left( 1-\frac{3\varepsilon }{2}+\frac{7}{4}%
\varepsilon ^{2}+\mathcal{O}(\varepsilon ^{3})\right) ,\qquad \bar{%
\varepsilon}=-\frac{1}{\bar{H}^{2}}\frac{\text{\textrm{d}}\bar{H}}{\text{%
\textrm{d}}\bar{t}}=3\varepsilon ^{2}-2\varepsilon ^{4}+\mathcal{O}%
(\varepsilon ^{5}),  \notag \\
\bar{\eta} &=&-2\varepsilon +\frac{13}{3}\varepsilon ^{2}+\mathcal{O}%
(\varepsilon ^{3}),\qquad H=\frac{m_{\phi }}{\sqrt{6\varepsilon }}\left( 1-%
\frac{\varepsilon }{12}+\frac{19\varepsilon ^{2}}{288}+\mathcal{O}%
(\varepsilon ^{3})\right) .  \label{epsa}
\end{eqnarray}%
The relations can be extended to arbitrary orders, if needed. We find $\bar{%
\eta}=\mathcal{O}(\bar{\varepsilon}^{1/2})$ and $\mathrm{d}^{n}\bar{%
\varepsilon}/$\textrm{d}$\bar{t}^{n}=\bar{H}^{n}\mathcal{O}(\bar{\varepsilon}%
^{(n+2)/2})$, which justifies the organization (\ref{assum}) of the
expansion around the de Sitter background in the inflaton framework. In
particular, inverting $\bar{\varepsilon}(\varepsilon )$ we get 
\begin{eqnarray}
\bar{H} &=&\frac{m_{\phi }}{2}\left( 1-\frac{\sqrt{3\bar{\varepsilon}}}{2}+%
\frac{7\bar{\varepsilon}}{12}-\frac{47\bar{\varepsilon}^{3/2}}{72\sqrt{3}}+%
\mathcal{O}(\bar{\varepsilon}^{2})\right) ,  \notag \\
\bar{\eta} &=&-2\sqrt{\frac{\bar{\varepsilon}}{3}}\left( 1-\frac{13}{6}\sqrt{%
\frac{\bar{\varepsilon}}{3}}-\frac{2\bar{\varepsilon}}{27}+\mathcal{O}(\bar{%
\varepsilon}^{3/2})\right) ,  \label{hbar} \\
-\bar{a}\bar{H}\tau &=&1+\bar{\varepsilon}+\frac{4\bar{\varepsilon}^{3/2}}{%
\sqrt{3}}+\frac{91}{9}\bar{\varepsilon}^{2}+\mathcal{O}(\bar{\varepsilon}%
^{5/2}).  \notag
\end{eqnarray}%
The last formula is derived from (\ref{ahHD}). Without passing through the
geometric framework, the relations (\ref{hbar}) can be worked out directly
in the inflaton framework by expanding the equations (\ref{frie}) around the
de Sitter background.

The map relating the fluctuations can be worked out from (\ref{weylmetric}).
The tensor modes $u$ and $v$ are clearly invariant,%
\begin{equation}
\bar{u}=\mathcal{W}_{0}\frac{a^{2}}{\bar{a}^{2}}u=u,\qquad \bar{v}=\mathcal{W%
}_{0}\frac{a^{2}}{\bar{a}^{2}}v=v,  \label{uv}
\end{equation}%
while the scalar fluctuations $\Psi $ and $\Phi $ transform as 
\begin{equation}
\quad \bar{\Psi}=\Psi +\frac{\Omega }{6(m_{\phi }^{2}+2\Upsilon )},\qquad 
\bar{\Phi}=\Phi -\frac{\Omega }{6(m_{\phi }^{2}+2\Upsilon )}.
\label{psitransf}
\end{equation}%
These formulas are written up to corrections of orders $\mathcal{O}(u\Omega
) $, $\mathcal{O}(v\Omega )$, $\mathcal{O}(\Psi \Omega )$ and $\mathcal{O}%
(\Phi \Omega )$, respectively. We can omit them for our purposes, since they
do not affect the quadratic action and the two-point functions. We recall
that the action is expanded around a solution of the equations of motion
(which is then expanded around the de Sitter metric -- which is not an exact
solution), so the linear terms in the fluctuations are absent. Switching
from one framework to the other, the corrections just mentioned affect the
cubic terms, but not the quadratic ones.

From (\ref{psitransf}) we derive the transformation of the curvature
perturbation $\mathcal{R}$. Observe that, given a scalar $Y=Y_{0}+\delta Y$,
where $\delta Y$ denotes the fluctuation around its background value $Y_{0}$%
, the combination 
\begin{equation}
\mathcal{R}_{Y}=\Psi +H\frac{\delta Y}{\dot{Y}_{0}}  \label{RY}
\end{equation}%
is invariant under infinitesimal time reparametrizations. If we choose $Y=%
\mathcal{W}$ and use the relations (\ref{relationsW}), we find, in the
geometric framework, 
\begin{equation}
\mathcal{R}_{\mathcal{W}}=\Psi -H\frac{m_{\phi }^{2}}{2\dot{\Upsilon}}\frac{%
\Omega }{3m_{\phi }^{2}}=\Psi -H\frac{\Omega }{6\dot{\Upsilon}}=\mathcal{R},
\label{r1}
\end{equation}%
the last equality following from (\ref{R}). Using (\ref{relationsW}), (\ref%
{mdsubs}) and (\ref{psitransf}) to rewrite this expression in the inflaton
framework, we obtain 
\begin{equation}
\mathcal{R}_{\mathcal{W}}=\bar{\Psi}+\bar{H}\left( \frac{\mathrm{d}\phi _{0}%
}{\mathrm{d}\bar{t}}\right) ^{-1}\delta \phi =\bar{\mathcal{R}},  \label{r2}
\end{equation}%
where $\phi _{0}$ is the background value of $\phi =\phi _{0}+\delta \phi $,
such that $\mathcal{W}_{0}=\mathrm{e}^{-\hat{\kappa}\phi _{0}}$. We recall
that in section (\ref{scalarocm}) the comoving gauge $\delta \phi =0$ was
used, so we just had $\bar{\mathcal{R}}=\bar{\Psi}$ there. Equations (\ref%
{r1}) and (\ref{r2}) prove that $\mathcal{R}=\bar{\mathcal{R}}$, so $%
\mathcal{R}$ is also invariant when we switch frameworks.

This fact, together with (\ref{uv}), ensures that the power spectra
calculated in the paper coincide in the two frameworks. We can use the
formulas (\ref{epsa}) to check it explicitly to the orders we have been
working with. Comparing (\ref{spec}) with (\ref{specHD}), (\ref{specHD2})
and (\ref{tiltHD}), we find%
\begin{equation*}
\bar{A}_{T}(\bar{H},\bar{\varepsilon})=A_{T}(H,\varepsilon ),\qquad \bar{n}%
_{T}(\bar{\varepsilon})=n_{T}(\varepsilon ).
\end{equation*}%
Finally, comparing (\ref{depsi}) and (\ref{ns}) with (\ref{des}), (\ref{ndHD}%
) and (\ref{specHDs}), it is easy to verify that%
\begin{equation*}
\bar{A}_{\mathcal{R}}(\bar{H},\bar{\varepsilon},\bar{\eta})=A_{\mathcal{R}%
}(H,\varepsilon ,\eta ),\qquad \bar{n}_{\mathcal{R}}(\bar{\varepsilon},\bar{%
\eta})=n_{\mathcal{R}}(\varepsilon ,\eta ).
\end{equation*}

\section{Superhorizon evolution}

\setcounter{equation}{0}\label{appe}

In this appendix we show that the curvature perturbation $\mathcal{R}$ can
be considered constant on superhorizon scales for adiabatic fluctuations of
the energy-momentum tensor, in particular after the metric fluctuations exit
the horizon and before they re-enter it. We start by showing this result in
the inflaton framework.

Consider the energy momentum tensor $T_{\mu \nu }$ with components%
\begin{equation*}
T_{00}=\rho (1+2\Phi )+\delta \rho ,\qquad T_{0i}=-\partial _{i}\delta
q,\qquad T_{ij}=a^{2}\delta _{ij}[p(1-2\Psi )+\delta p]+\left( \partial
_{i}\partial _{j}-\frac{\triangle }{3}\delta _{ij}\right) \delta \Pi ,
\end{equation*}%
where $\delta \rho $, $\delta q$, $\delta p$ and $\delta \Pi $ are its
scalar fluctuations around the background. The gauge invariant curvature
perturbation is 
\begin{equation}
\mathcal{R}=\Psi -\frac{H}{\rho +p}(\delta q+pB).  \label{curv}
\end{equation}%
The unprojected equations derived from the action (\ref{sqgeq}) for the
metric (\ref{mets}) in the Newton gauge ($B=0$) read 
\begin{eqnarray}
2\dot{\Psi}+2H\Phi -\frac{2\triangle \dot{W}}{3m_{\chi }^{2}a^{2}}+8\pi
G\delta q &=&0,\qquad  \notag \\
\Phi -\Psi +\frac{1}{m_{\chi }^{2}}\left( \ddot{W}+H\dot{W}-\frac{\triangle W%
}{3a^{2}}\right) +8\pi G\delta \Pi &=&0,  \notag \\
6H\dot{\Psi}-\frac{2\triangle \Psi }{a^{2}}+\frac{2\triangle ^{2}W}{3m_{\chi
}^{2}a^{4}}+8\pi G(\delta \rho +2\rho \Phi ) &=&0,  \notag \\
\ddot{\Psi}+H(3\dot{\Psi}+\dot{\Phi})+2\Phi \dot{H}+3H^{2}\Phi +\frac{%
\triangle (\Phi -\Psi )}{3a^{2}}-\frac{\triangle ^{2}W}{9m_{\chi }^{2}a^{4}}%
-4\pi G\delta p &=&0,  \label{eqque}
\end{eqnarray}%
where $W=\Psi +\Phi $ and the contributions of the scalar field $\phi $ are
moved into $T_{\mu \nu }$. It is possible to show that formulas (\ref{eqque}%
), together with the Friedmann equations%
\begin{equation*}
H^{2}=\frac{8\pi G}{3}\rho ,\qquad \dot{H}+\frac{3}{2}H^{2}=-4\pi Gp,
\end{equation*}%
imply the equation%
\begin{equation}
\frac{\dot{H}}{H}\dot{\mathcal{R}}=-4\pi G\left( \delta p-\frac{\dot{p}}{%
\dot{\rho}}\delta \rho \right) -\frac{\triangle }{a^{2}}\left[ \frac{\dot{p}%
}{\dot{\rho}}\left( \Psi -\frac{\triangle W}{3m_{\chi }^{2}a^{2}}-\frac{H%
\dot{W}}{m_{\chi }^{2}}\right) -\frac{\dot{H}\dot{W}}{3m_{\chi }^{2}H}+\frac{%
8\pi G}{3}\delta \Pi \right] .  \label{rdot}
\end{equation}

Thus, for adiabatic fluctuations 
\begin{equation*}
\delta p=\frac{\dot{p}}{\dot{\rho}}\delta \rho
\end{equation*}%
and on superhorizon scales $k/(aH)\ll 1$, the scalar $\mathcal{R}$ is
practically constant. Since the property holds for the whole set of
solutions of the unprojected equations, it also holds for the projected
ones. Note that after the end of inflation $\varepsilon $ is no longer
small, so the $\dot{H}$ factor in front $\dot{\mathcal{R}}$ in (\ref{rdot})
is not a source of trouble.

In the geometric framework we reach the same conclusions. It is sufficient
to work with the action (\ref{SQGmix}) and note that the only difference
with respect to the formulas just written is a redefinition of $T_{\mu \nu }$%
, brought by the variation of the terms containing $\varphi $ with respect
to the metric. Since we are considering only scalar quantities here, this is
just a redefinition of $\rho $, $p$ and the fluctuations $\delta \rho $, $%
\delta q$, $\delta p$ and $\delta \Pi $. Observe that we may need a
nontrivial $\delta \Pi $ for this redefinition, which is the reason why we
kept it nonzero in the derivation above.

\end{document}